\documentclass[11pt]{article}
\usepackage{epsfig}
\usepackage{axodraw}
\def\be{\begin{equation}}
\def\ee{\end{equation}}
\def\bc{\begin{center}}
\def\ec{\end{center}}
\def\bea{\begin{eqnarray}}
\def\eea{\end{eqnarray}}

\def\nn{\nonumber}

\def\ov{\overline}
\def\hlf{\frac{1}{2}}

\def\as{\alpha_s}
\def\at{\alpha_t}
\def\ab{\alpha_b}
\def\sq2{\sqrt{2}}
\def\ths{\bar{\theta}_{\tilde{t}}}
\def\thz{\theta_{\tilde{t}}}
\def\mix{\widetilde{X}}
\def\mixt{X}
\def\c2t2{c_{2\bar{\theta}}^{\,2}}

\def\s2t{s_{2\theta}}
\def\c2t{c_{2\theta}}
\def\ct{c_{\theta}}
\def\st{s_{\theta}}

\def\mgl{m_{\tilde{g}}}
\def\msqu{m_{\tilde{t}_1}^2}
\def\msqd{m_{\tilde{t}_2}^2}

\def\mbl{m^2_{\tilde{b}_L}}
\def\mtu{m_{\tilde{t}_1}}
\def\mtd{m_{\tilde{t}_2}}
\def\cpp{c_{\varphi\tilde{\varphi}}}
\def\v2lp{V^{\rm 2loop}}

\def\mt{m_t}
\def\sb{s_\beta}
\def\cb{c_\beta}
\def\simlt{\stackrel{<}{{}_\sim}}
\def\simgt{\stackrel{>}{{}_\sim}}
\newcommand{\smallz}{{\scriptscriptstyle Z}} %
\newcommand{\mz}{m_\smallz}
\newcommand{\ma}{m_{\scriptscriptstyle A}}
\newcommand{\mw}{m_{\scriptscriptstyle W}}
\newcommand{\mabar}{\overline{m}_{\scriptscriptstyle A}}
\newcommand{\mh}{m_h}
\newcommand{\mH}{m_{\scriptscriptstyle H}}
\newcommand{\mHbar}{\overline{m}_{\scriptscriptstyle H}}

\def\diff{\msqu-\msqd}

\def\mhg{\mu}
\def\li2{{\rm Li}_2}
\def\T{M_S^2}
\def\Tq{M_S^4}
\def\t{\mt^2}
\def\A0{\ma^2}
\def\Aq{\ma^4}
\def\logt{\ln\frac{\t}{Q^2}}
\def\logT{\ln\frac{\T}{Q^2}}
\def\logTt{\ln\frac{\T}{\t}}
\def\logA0{\ln\frac{\,\A0}{Q^2}}
\def\logAt{\ln\frac{\,\A0}{\t}}
\def\logAT{\ln\frac{\,\A0}{\T}}

\def\logqA0{\ln^2\frac{\A0}{Q^2}}

\catcode`@=11
\def\marginnote#1{}
\newcount\hour
\newcount\minute
\newtoks\amorpm
\hour=\time\divide\hour by60
\minute=\time{\multiply\hour by60 \global\advance\minute by-\hour}
\edef\standardtime{{\ifnum\hour<12 \global\amorpm={am}%
        \else\global\amorpm={pm}\advance\hour by-12 \fi
        \ifnum\hour=0 \hour=12 \fi
        \number\hour:\ifnum\minute<10 0\fi\number\minute\the\amorpm}}
\edef\militarytime{\number\hour:\ifnum\minute<10 0\fi\number\minute}
\def\draftlabel#1{{\@bsphack\if@filesw {\let\thepage\relax
   \xdef\@gtempa{\write\@auxout{\string
      \newlabel{#1}{{\@currentlabel}{\thepage}}}}}\@gtempa
   \if@nobreak \ifvmode\nobreak\fi\fi\fi\@esphack}
        \gdef\@eqnlabel{#1}}
\def\@eqnlabel{}
\def\@vacuum{}
\def\draftmarginnote#1{\marginpar{\raggedright\scriptsize\tt#1}}
\def\draft{\oddsidemargin 0.0truein
        \def\@oddfoot{\sl preliminary draft \hfil
        \rm\thepage\hfil\sl\today\quad\militarytime}
        \let\@evenfoot\@oddfoot \overfullrule 3pt
        \let\label=\draftlabel
        \let\marginnote=\draftmarginnote
   \def\@eqnnum{(\theequation)\rlap{\kern\marginparsep\tt\@eqnlabel}%
\global\let\@eqnlabel\@vacuum}  }
\catcode`@=12
\newenvironment{appendletterA}
 {
  \typeout{ Starting Appendix \thesection }
  \setcounter{section}{0}
  \setcounter{equation}{0}
  \renewcommand{\theequation}{A\arabic{equation}}
 }{
  \typeout{Appendix done}
 }

\begin{document}
\thispagestyle{empty}
\begin{center}
\hfill{DFPD-01/TH/40}\\
\hfill{RM3-TH/01-14} \\
\hfill{ROME1-1327/01} \\
\vspace{1.7cm}
\bc
{\LARGE\bf On the ${\cal O} (\at^2)$ two--loop corrections to the}
\ec
\bc
{\LARGE\bf neutral Higgs boson masses in the MSSM}
\ec
\vspace{1.4cm}
{\Large \sc A.~Brignole~$^{a}$, G.~Degrassi~$^{a,b}$,
P.~Slavich~$^{a}$ and F.~Zwirner~$^{c}$}\\

\vspace{1.2cm}

${}^a$
{\em 
Dipartimento di Fisica `G.~Galilei', Universit\`a di Padova and
\\ 
INFN, Sezione di Padova, Via Marzolo~8, I-35131 Padua, Italy
}
\vspace{.3cm}

${}^b$
{\em 
Dipartimento di Fisica, Universit\`a di Roma III
\\
Via della Vasca Navale~84, I-00146 Rome, Italy
}
\vspace{.3cm}

${}^c$
{\em 
Dipartimento di Fisica, Universit\`a di Roma `La Sapienza' and
\\
INFN, Sezione di Roma, P.le Aldo Moro~2, I-00185 Rome, Italy}
\end{center}

\vspace{0.8cm}

\centerline{\bf Abstract}
\vspace{2 mm}
\begin{quote}\small
We compute the ${\cal O}(\at^2)$ two--loop corrections to the neutral
CP--even Higgs boson mass matrix in the Minimal Supersymmetric
Standard Model, for arbitrary values of $\ma$ and of the parameters in
the stop sector, in the effective potential approach. In a large
region of parameter space these corrections are sizeable, increasing
the prediction for $m_h$ by several GeV. We present explicit
analytical formulae for a simplified case. We discuss the inclusion of
momentum--dependent corrections and some possible ways of assigning
the input parameters.
\end{quote}

\vfill
\newpage
\setcounter{equation}{0}
\setcounter{footnote}{0}
\vskip2truecm
%
%%%%%%%%%%%%%%%%%%%%%%%%%%%%%%%%%%%%%%%%%%%%%%%%%%%%%%%%%%%%%%%%%%
%
\section{Introduction}
\label{section:1}
The Minimal Supersymmetric extension of the Standard Model, or MSSM
(for a recent review and references, see e.g. Ref.~\cite{mssm}) is today a
paradigm for possible new physics at the weak scale, supported by many
theoretical and phenomenological considerations. The predictive power
of the MSSM, however, is strongly limited by our ignorance of the
mechanism for supersymmetry breaking in the underlying fundamental
theory. In particular, the spectrum of supersymmetric particles, with
its strong dependence on many soft supersymmetry--breaking parameters,
is essentially unconstrained. This is not true for the MSSM Higgs
sector: the dimensionless Higgs couplings in the scalar potential are
controlled by supersymmetric Ward identities that receive only finite
quantum corrections from the soft breaking of supersymmetry.  As a
result, the masses and the couplings of the MSSM Higgs bosons,
$(h,H,A,H^\pm)$, can be expressed, at the classical level, in terms of
the gauge boson masses $(\mw,\mz)$ and only two additional parameters:
for example, the mass of the neutral CP--odd state, $\ma$, and the
ratio of the two Higgs vacuum expectation values (VEVs), $\tan
\beta$. The number of independent predictions in the Higgs sector does
survive at the quantum level, but radiative corrections induce finite
modifications of such predictions, and bring along additional
dependences on the remaining MSSM parameters, associated with the
virtual particles in the loops. This is the reason why an impressive
theoretical effort has been devoted to the computation of the
radiative corrections to the MSSM Higgs sector.

In this paper we address once more the crucial issue of the
computation of the neutral CP--even Higgs boson masses. Soon after the
first calculations of the dominant one--loop effects, associated with
top and stop loops \cite{first}, the full one--loop computation became
available \cite{ab,onefull,bmpz}, and resummations of leading and
next--to--leading logarithms were included \cite{rge} via appropriate
renormalization group equations (RGEs). We are now at the stage in
which the computation of the most important two--loop effects is being
completed. In analogy with the one--loop case, the most important
two--loop effects are generated by strong and top--Yukawa corrections
to the one--loop diagrams involving the top quark and its
supersymmetric partners, i.e. they are ${\cal O}(\at \as)$ and ${\cal
O}(\at^2)$, respectively.

The first genuine two--loop computation \cite{hh} considered ${\cal
O}(\at \as)$ and ${\cal O} (\at^2)$ corrections to $m_h$, and was
restricted to the limiting case of $\tan \beta \to \infty$ and
negligible stop mixing.

At ${\cal O}(\at \as)$, the result of \cite{hh} was generalized by
three groups, all working in the limit of vanishing external momentum,
i.e. in the effective potential approximation (EPA).  The diagrammatic
computation of \cite{hollik} is valid for arbitrary values of $\ma$
and of the stop parameters, but in such general case its complete
formulae are rather lengthy and available only in the form of a
computer code. The computation of \cite{ez1} is applicable to the
corrections to $m_h$, in the limit $\ma \gg \mz$, and its results are
available in analytic form under further restrictions.  The
computation of \cite{DSZ} has recently provided simple analytical
formulae for the two--loop corrected mass matrix of the neutral
CP--even Higgs sector, for arbitrary values of $\ma$ and of the stop
parameters.

The inclusion of the ${\cal O} (\at^2)$ corrections, which can be of
comparable numerical importance, is still at an earlier stage.  After
\cite{hh}, the only genuine two--loop computation so far is the one of
\cite{ez2}, which, as the one of \cite{ez1}, is applicable to the
corrections to $m_h$, in the limit $\ma \gg \mz$. Moreover, the full
results of \cite{ez2} are available only in numerical form, but simple
analytical formulae are provided under the additional assumptions of
large and universal soft stop masses, $m_Q^2 = m_U^2 \gg m_t^2$. In
the present paper we go one step further, and compute the ${\cal O}
(\at^2)$ two--loop corrections to the mass matrix of the neutral
CP--even Higgs bosons. We still work in the EPA, but our calculation
is valid for arbitrary values of $\ma$ and of the stop parameters.

Our paper is organized as follows. After this introduction, Section~2
recalls some general features of the calculation of the neutral Higgs
boson masses in the MSSM. In particular, we describe the connection 
of the EPA with the full calculation, and an `improved' perturbative
determination of the masses from the full momentum--dependent
two--point function.  Section~3 describes the main features of our
two--loop calculation of the ${\cal O} (\at^2)$ contributions.
Explicit analytical formulae are displayed for the simple case of
large and universal soft stop masses, but arbitrary values of all the
remaining Higgs and stop parameters.  Section~4 discusses in detail
some possible ways of assigning the input parameters, and the relation
between the $\ov{\rm DR}$ scheme and on--shell (OS) renormalization
schemes, identifying a version of the latter that we find particularly
convenient.  Section~5 presents some quantitative evaluation of the
${\cal O} (\at^2)$ corrections to $m_h$ and $\mH$, and numerical
comparisons with the previously available results, for a number of
representative cases. We find that the ${\cal O}(\at^2)$ corrections
can be sizeable in a large region of parameter space, increasing the
prediction for $m_h$ by several GeV.  In particular, for large stop
mixing most of these corrections are genuine two--loop effects, which
cannot be accounted for by standard renormalization--group
improvements. In the concluding section we summarize our
results and comment on possible extensions. Technical details, such 
as the explicit formulae that are needed for the transition from the
$\ov{\rm DR}$ scheme to our implementation of the OS scheme, are
confined to Appendix A. We find that our general analytical result
is too long even for an appendix, thus we make it available, upon
request, in the form of a computer code~\footnote{E--mail: {\tt
pietro.slavich@pd.infn.it}.}.
%
%%%%%%%%%%%%%%%%%%%%%%%%%%%%%%%%%%%%%%%%%%%%%%%%%%%%%%%%%%%%%%%%%%
%
\section{General results}
\label{section:2}

We review here, along the lines of \cite{ab}, some general results
that apply to the calculation of the MSSM neutral Higgs boson masses,
in the CP--conserving case, at every order in the different coupling
constants and in the loop expansion.

In the EPA, the CP--odd and CP--even mass matrices are identified
with the second derivatives of the effective potential, $V_{\rm eff} 
= V_0 + V$, evaluated at its minimum:
\be
\label{defmat}
\left({\cal M}^2_P\right)_{ij}^{\rm eff} \equiv
\left. \frac{\partial^2  V_{\rm eff}}{\partial P_i \partial P_j}
\right|_{\rm min} \, , 
\hspace{1cm}
\left({\cal M}^2_S\right)_{ij}^{\rm eff} \equiv
\left. \frac{\partial^2  V_{\rm eff}}{\partial S_i \partial S_j}
\right|_{\rm min} \, ,
\hspace{1cm}
(i,j=1,2) \, ,
\ee
where we have decomposed the Higgs fields into their VEVs plus their 
CP--even and CP--odd fluctuations as follows:
\be 
\label{expansion}
H_1^0 \equiv {v_1 + S_1 + i \, P_1 \over \sq2} \, ,
\;\;\;\;\;\;\;
H_2^0 \equiv {v_2 + S_2 + i \, P_2 \over \sq2} \, .
\ee
It is understood that $V_{\rm eff}$ is expressed in terms of
renormalized fields and parameters, in the Landau gauge and in 
the $\ov{\rm DR}$ renormalization scheme~\footnote{In our 
decomposition, $V_0$ has the same functional form of the
tree--level potential, whereas $V$ contains the residual
loop corrections. The gauge choice will not be relevant 
for the two--loop calculations of the present paper.}. 
In particular:
\be
\left( {\cal M}^2_S \right)^{\rm eff}  = 
\left( {\cal M}^2_S \right)^{0, \, {\rm eff}} 
+
\left(\Delta{\cal M}^2_S\right)^{\rm eff}  \, ,
\label{effmatrix}
\ee
where
\be
\left( {\cal M}^2_S \right)^{0, \, {\rm eff}}
= 
\left(\begin{array}{cc}
\ov m_\smallz^2 \, \cb^2 + \ov \ma^2 \,\sb^2 
& -\left(\ov m_\smallz^2 + \ov \ma^2 \right) \sb\, \cb \\ 
-\left(\ov m_\smallz^2 + \ov \ma^2\right) \sb\, \cb 
& \ov m_\smallz^2 \, \sb^2 + \ov \ma^2 \,\cb^2 
\end{array}\right) 
\label{mpma} 
\, ,
\ee
\be
\left(\Delta{\cal M}^2_S\right)_{ij}^{\rm eff}  = 
-(-1)^{i+j}\left.
\frac{\partial^2 V}{\partial P_i \partial P_j}\right|_{\rm min}
+\left.\frac{\partial^2 V}{\partial S_i \partial S_j}
\right|_{\rm min} \, .
\label{Dms}
\ee

In Eq.~(\ref{mpma}), $\mabar^2$ is the mass of the CP--odd state
in the EPA, i.e. the non--vanishing eigenvalue of $\left( {\cal M}^2_P 
\right)^{\rm eff}$. Similarly, $\ov{m}_\smallz^2 = (g^2 + g'^2) \, 
v^2/4$ is just the $\ov{\rm DR}$ mass for the $Z$ boson~\footnote{This
choice is not imposed by the EPA, but is rather a matter of convenience.
Alternatively, we could include in $\ov{m}_\smallz^2$ the self--energy
corrections evaluated at vanishing external momentum.}, where $v^2 = 
v_1^2 + v_2^2$, and $v_1$ and $v_2$ are the $\ov{\rm DR}$ VEVs computed 
by minimizing $V_{\rm eff}$. Finally, $\sb \equiv \sin\beta$ and 
$\cb \equiv \cos\beta$, where $\tan\beta = v_2/v_1$.

Since $V_{\rm eff}$ generates one--particle--irreducible (1PI) Green's
functions at vanishing external momentum, it is clear that the EPA
neglects some momentum--dependent effects in the Higgs self--energies,
and the entire finite contribution to the $Z$ self--energy. The
complete computation requires the full two--point
function~\footnote{We absorb in our $\Gamma$'s the usual $i$ factor
and consider only their real part.  The same conventions apply then to
the self--energies.}  $\Gamma_S(p^2)$, which takes the form of a $2
\times 2$ matrix in the space $(S_1,S_2)$, and the analogous
two--point functions for the neutral CP--odd spin--0 bosons and gauge
bosons. Using the standard decomposition of the two--point function,
\be
\Gamma_S \, (p^2) = p^2  - {\cal M}^2_S(p^2) \, ,
\label{defgamma}
\ee
the physical masses $(m_h^2,\mH^2)$ are the two solutions of 
the equation
\be 
{\rm det} \left[  p^2  - {\cal M}^2_S(p^2) \right] = 0 \, . 
\label{3.49a}
\ee
At any loop order, a full diagrammatic calculation relates 
these two masses with the physical masses of the neutral 
CP--odd Higgs boson ($\ma^2$) and of the $Z$ vector boson 
($\mz^2$), defined in a similar way. 

To make contact between the complete calculation and the EPA, we need
the relations between $[{\cal M}^2_S(p^2), \ma^2,\mz^2]$ on the one
side, and $[({\cal M}^2_S)^{\rm eff}, \mabar^2,\ov{m}_{\smallz}^2]$ on
the other side. The transverse $ZZ$ propagator and the $AA$ propagator
have the form $[p^2 - \ov{m}_{\smallz}^2 +\Pi_{ZZ}(p^2)]^{-1}$ and
$[p^2 - \mabar^2 + \delta \Pi_{AA}(p^2)]^{-1}$, respectively. Thus we
have
\bea 
\ov m_\smallz^2 &=& \mz^2 + \Pi_{ZZ}(\mz^2) \, ,
\label{mz} \\
\ov \ma ^2 &=&  \ma^2 + \delta  \Pi_{AA}(\ma^2) \, . 
\label{ma}
\eea 
It is understood here that the ($\ov{\rm DR}$--renormalized) two point
functions $\Pi_{ZZ}(p^2)$ and $\delta\Pi_{AA}(p^2)$ include both 1PI
and non--1PI (mixing) terms. Indeed, beyond the tree level the
transverse $Z$ boson mixes with the photon, whereas the the field
$A=\sin \beta \, P_1 + \cos \beta \, P_2$ mixes with the Goldstone
boson $G= -\cos \beta \, P_1 + \sin \beta \, P_2$ and with the
longitudinal $Z$ boson. In the gaugeless limit, for instance, $\delta
\Pi_{AA}(p^2)$ reads
\be
\delta  \Pi_{AA}(p^2) = \Delta\Pi_{AA}(p^2) - 
\frac{[\Delta\Pi_{AG}(p^2)]^2}{p^2 + \Delta\Pi_{GG}(p^2)} \, ,
\ee
where, for a generic self--energy $\Pi_{ab}(p^2)$, we define
\be
\Delta \Pi_{ab}(p^2) \equiv \Pi_{ab}(p^2) - \Pi_{ab}(0) 
\, .
\label{deltaA}
\ee
For the two--point function of the CP--even Higgs fields,
we can write 
\be
{\cal M}^2_S\,(p^2) =
\left( {\cal M}^2_S \right)^{0} + 
\left( \Delta{\cal M}^2_S \right)^{\rm eff} + 
\left( \Delta{\cal M}^2_S\right)^{p^2} \, ,
\label{defM} 
\ee
where $({\cal M}^2_S)^{0}$ has the same form of $({\cal M}^2_S)^{0, \,
{\rm eff}}$ in Eq.~(\ref{mpma}), with the only difference that the EPA
masses $(\mabar^2,\ov{m}_\smallz^2)$ are replaced by the physical
masses $(\ma^2,\mz^2)$. The matrix $(\Delta{\cal M}^2_S)^{\rm eff}$
was given in Eq.~(\ref{Dms}). Finally, the matrix $(\Delta{\cal
M}^2_S)^{p^2}$ has entries
\bea
\left( \Delta{\cal M}^2_S\right)^{p^2}_{11} &=&
\Pi_{ZZ}(\mz^2) \, \cb^2 +  \delta  \Pi_{AA}(\ma^2) \,\sb^2 
- \Delta \Pi_{S_1 S_1}\,(p^2) \, , \label{defMp11} \\ 
\left( \Delta{\cal M}^2_S\right)^{p^2}_{12} &=&
-\left[ \Pi_{ZZ}(\mz^2) + \delta   \Pi_{AA}(\ma^2) \right] \sb\, \cb 
- \Delta \Pi_{S_1 S_2}\,(p^2) \, , \label{defMp12} \\ 
\left( \Delta{\cal M}^2_S\right)^{p^2}_{22} &=&
  \Pi_{ZZ}(\mz^2) \, \sb^2 + \delta  \Pi_{AA}(\ma^2) \,\cb^2 
- \Delta \Pi_{S_2 S_2}\,(p^2) \, .
\label{defMp22} 
\eea
This establishes the desired correspondence between the EPA and the
full computation.

We now describe how to perform an `improved' perturbative
determination of the neutral CP--even Higgs masses (and of the
associated mixing angle), starting from Eq.~(\ref{3.49a}). Of course,
this equation could be solved numerically, at the cost of some
computer time. Our solution has the advantage of producing explicit
and accurate analytical formulae, which can facilitate a number of
checks and considerably speed up the numerical evaluation of ($m_h^2,
\mH^2$), for any given set of input parameters. The latter feature is 
welcome for the practical implementation of our results in computer
codes.  We have explicitly checked that, for a wide range of parameter
choices, our perturbative solutions of Eq.~(\ref{3.49a}) agree with
numerical solutions within $10^{-3}$~GeV.

The matrix ${\cal M}^2_S \,(p^2)$ can be conventionally decomposed 
into the sum of a $p^2$--independent `unperturbed' piece and a 
$p^2$--dependent `perturbation': 
\be
{\cal M}^2_S \, (p^2) = \ov{\cal M}^2 + \Delta {\cal M}^2 (p^2) \, .
\label{3.50a}
\ee
At the lowest order, i.e.~neglecting  $\Delta {\cal M}^2 (p^2)$, 
Eq.~(\ref{3.49a}) reduces to the diagonalization of $\ov{\cal M}^2$.
This gives the eigenvalues
\be
\ov{m}^2_{{\scriptscriptstyle H},h} = \frac12 \left[ \, \ov{\cal M}^2_{11}+ \ov{\cal M}^2_{22} 
   \pm \sqrt{\left( \ov{\cal M}^2_{11}- \ov{\cal M}^2_{22} \right)^2 +
   4  \left( \ov{\cal M}^2_{12} \right)^2 } \, \right] \, ,
\label{3.51a}
\ee
and the eigenstates
\be
\left( \begin{array}{c} H \\ h \end{array} \right) =
\left( \begin{array}{cc} \cos \ov{\alpha} & \sin \ov{\alpha} \\
-\sin \ov{\alpha} & \cos \ov{\alpha}  \end{array} \right)\,
\left( \begin{array}{c} S_1 \\ S_2 \end{array} \right) \, ,
\label{3.53a}
\ee
where the mixing angle $\ov{\alpha}$ is determined by
\be
\sin \, 2 \ov{\alpha}  = 
\frac{2\,\ov{\cal M}^2_{12}}{\mHbar^2 - \ov{m}_h^2} \, ,
\;\;\;\;\;
\cos \, 2 \ov{\alpha}  = 
\frac{\ov{\cal M}^2_{11}- \ov{\cal M}^2_{22}}{\mHbar^2 
- \ov{m}_h^2} \, ,
\;\;\;\;\;
(-\pi/2 < {\ov \alpha} < \pi/2) \, .
\label{3.54a}  
\ee
At first order in $\Delta {\cal M}^2 (p^2)$, we can write 
for the two eigenvalues
\bea
[\mh^2]^1 
& = &
\ov{m}_h^2 
+ \sin^2 \ov{\alpha} \, \Delta {\cal M}^2_{11} (\ov{m}_h^2)  
- \sin 2 \ov{\alpha} \, \Delta {\cal M}^2_{12} (\ov{m}_h^2) 
+ \cos^2 \ov{\alpha} \, \Delta {\cal M}^2_{22} (\ov{m}_h^2) 
\, ,
\label{mhfirst}\\[0.3cm]
[\mH^2]^1 &  = & 
\mHbar^2 
+ \cos^2 \ov{\alpha} \, \Delta {\cal M}^2_{11} (\mHbar^2) 
+ \sin 2 \ov{\alpha} \, \Delta {\cal M}^2_{12} (\mHbar^2)
+ \sin^2 \ov{\alpha} \, \Delta {\cal M}^2_{22} (\mHbar^2)
\, .
\label{mHfirst}
\eea
At the second order:
\bea
\mh^2 & = & 
\ov{m}_h^2 
+ \sin^2 \ov{\alpha} \, \Delta {\cal M}^2_{11} ([\mh^2]^1)  
- \sin 2 \ov{\alpha} \, \Delta {\cal M}^2_{12} ([\mh^2]^1) 
+ \cos^2 \ov{\alpha} \, \Delta {\cal M}^2_{22} ([\mh^2]^1) 
\nn \\
& - & \frac{1}{\mHbar^2 - \ov{m}_h^2} 
\left\{ \frac{\sin 2 \ov{\alpha} }{2} \, 
\left[ \Delta {\cal M}^2_{22} (\ov{m}_h^2)-\Delta {\cal M}^2_{11} 
(\ov{m}_h^2) \right] + \cos 2 \ov{\alpha} \,\Delta {\cal M}^2_{12} 
(\ov{m}_h^2) \right\}^2 
\, ,
\label{mhcorr}\\[0.3cm]
\mH^2 & = & 
\mHbar^2 
+ \cos^2 \ov{\alpha} \, \Delta {\cal M}^2_{11} ([\mH^2]^1) 
+ \sin 2 \ov{\alpha} \, \Delta {\cal M}^2_{12} ([\mH^2]^1) 
+ \sin^2 \ov{\alpha} \, \Delta {\cal M}^2_{22} ([\mH^2]^1) 
\nn \\
& + & \frac{1}{\mHbar^2 - \ov{m}_h^2} 
\left\{ \frac{ \sin 2 \ov{\alpha}}{2}  \, 
\left[ \Delta {\cal M}^2_{22} (\mHbar^2)
- \Delta {\cal M}^2_{11} (\mHbar^2) 
\right]  + \cos 2 \ov{\alpha}  \, 
\Delta {\cal M}^2_{12} (\mHbar^2) \right\}^2 
\, .
\label{mHcorr}
\eea

As discussed in \cite{ab}, the decomposition of Eq.~(\ref{3.50a}) is
not unique. It is convenient to specify it in a way that improves, as
much as possible, the accuracy of the perturbative solutions of
Eq.~(\ref{3.49a}), given by Eqs.~(\ref{mhcorr}) and (\ref{mHcorr}),
and allows to absorb an important part of the corrections to the Higgs
interaction vertices into the definition of the renormalized mixing
angle $\ov{\alpha}$. The natural choice $\ov{\cal M}^2 = ({\cal
M}^2_S)^{0}$ and $\Delta {\cal M}^2 (p^2) = (\Delta {\cal M}^2_S)^{\rm
eff} + (\Delta{\cal M}^2_S)^{p^2}$ can work well for $\ma \gg \mz$,
while it is known~\cite{ab} not to be very effective for $\ma \sim
\mz$.  A choice that leads to an improved convergence, and will be
adopted in our numerical evaluation, is instead
\be
\ov{\cal M}^2 = \left( {\cal M}^2_S \right)^{0} + 
\left( \Delta{\cal M}^2_S \right)^{\rm eff} \, ,
\;\;\;\;\;
\Delta {\cal M}^2 (p^2) = 
\left( \Delta{\cal M}^2_S\right)^{p^2} \, ,
\label{mbar} 
\ee
although other choices can be equally valid. 

Once the decomposition of ${\cal M}^2_S \,(p^2)$ has been specified,
the perturbative solutions described above can be written more
explicitly, taking also into account the conventional loop expansion
of the self--energies ($\Pi=\Pi^{(1)}+\Pi^{(2)}+\ldots $).  Consider
the decomposition specified by (\ref{mbar}).  Then the two--loop
corrected Higgs masses can be written as
\bea
m_h^2 & = &  \ov{m}_h^2  + \delta^{(1)} m_h^2 + \delta^{(2)} m_h^2 \, ,
\label{mh012}
\\
m_H^2 & = &  \mHbar^2 + \delta^{(1)} m_H^2 + \delta^{(2)} m_H^2 \, ,
\label{mH012}
\eea
where the two--loop corrected masses obtained in the EPA, $(\ov{m}_h^2,
\mHbar^2)$, are to be supplemented by the corrections that stem from
$(\Delta{\cal M}^2_S)^{p^2}$.  At first order we have \cite{ab}:
\bea
\delta^{(1)} \mh^2 
& = &
 \sin^2 (\beta +\ov{\alpha}) \, \Pi_{ZZ}^{(1)}(\mz^2) 
+ \cos^2 (\beta -\ov{\alpha}) \, \Delta \Pi_{AA}^{(1)}(\ma^2)  
- \Delta \Pi_{hh}^{(1)} (\ov{m}_h^2)
\, , 
\label{mhone}\\[0.3cm]
\delta^{(1)} \mH^2
&  = & 
 \cos^2 (\beta +\ov{\alpha}) \, \Pi_{ZZ}^{(1)}(\mz^2) 
+ \sin^2 (\beta -\ov{\alpha}) \, \Delta  \Pi_{AA}^{(1)}(\ma^2)  
- \Delta \Pi_{HH}^{(1)} (\mHbar^2)
\, .
\label{mHone}
\eea
At the next order:
\bea
\delta^{(2)} \mh^2 
& = & 
  \sin^2 (\beta +\ov{\alpha}) \, \Pi_{ZZ}^{(2)}(\mz^2) 
+ \cos^2 (\beta -\ov{\alpha}) \, \delta \Pi_{AA}^{(2)}(\ma^2)  
- \Delta \Pi_{hh}^{(2)} (\ov{m}_h^2)
\nn \\[0.1cm]
& - & 
\left[ \Delta \Pi_{hh}^{(1)} (\ov{m}_h^2  + \delta^{(1)} \mh^2 )
- \Delta \Pi_{hh}^{(1)} (\ov{m}_h^2) \right] 
\nn \\[0.1cm]
& - &  \! \frac{1}{\mHbar^2 \! - \! \ov{m}_h^2} \! 
\left[ \frac{ \sin 2(\beta \! + \! \ov{\alpha})}{2}  
\, \Pi_{ZZ}^{(1)}(\mz^2) \! + \! \frac{ \sin 2(\beta
\! - \! \ov{\alpha})}{2} \Delta \Pi_{AA}^{(1)}(\ma^2) 
\! + \! \Delta \Pi_{Hh}^{(1)} (\ov{m}_h^2) \right]^2 \! ,
\label{mhsec}\\[0.3cm]
\delta^{(2)} \mH^2
&  = & 
 \cos^2 (\beta +\ov{\alpha}) \, \Pi_{ZZ}^{(2)}(\mz^2) 
+ \sin^2 (\beta -\ov{\alpha}) \, \delta  \Pi_{AA}^{(2)}(\ma^2)  
- \Delta \Pi_{HH}^{(2)} (\mHbar^2)
\nn \\[0.1cm]
& - & 
\left[ \Delta \Pi_{HH}^{(1)} (\mHbar^2 + \delta^{(1)} \mH^2 )
- \Delta \Pi_{HH}^{(1)} (\mHbar^2) \right] 
\nn \\[0.1cm]
& + & \! \frac{1}{\mHbar^2 \! - \! \ov{m}_h^2} \! 
\left[ \frac{ \sin 2(\beta \! + \! \ov{\alpha})}{2}  
\, \Pi_{ZZ}^{(1)}(\mz^2) \! + \! \frac{ \sin 2(\beta 
\! - \! \ov{\alpha})}{2} \Delta \Pi_{AA}^{(1)}(\ma^2) 
\! + \! \Delta \Pi_{Hh}^{(1)} (\mHbar^2) \right]^2 \! .
\label{mHsec}
\eea
The corrections $(\delta^{(1)} \mh^2, \delta^{(1)} \mH^2)$ involve
one--loop self--energies that are known and should be included to
correct the EPA results \cite{ab}. On the other hand, the two--loop
effects in $(\delta^{(2)} \mh^2, \delta^{(2)} \mH^2)$ arise either
from products (or iterations) of one--loop self--energies, or from 1PI
two--loop contributions in $(\Pi_{ZZ}^{(2)}, \delta\Pi_{AA}^{(2)},
\Delta\Pi_{hh}^{(2)}, \Delta\Pi_{HH}^{(2)})$, which have not been 
computed so far. However, the effect of $\delta^{(2)} \mh^2$ 
is always subleading, and the same is true for  
$\delta^{(2)} \mH^2$ when $m_A$ is not large.

To qualify the latter statement, we recall that the dominant two--loop
corrections to $(\mh^2, \mH^2)$ are ${\cal O}(\at \as m_t^2)$ and
${\cal O}(\at^2 m_t^2)$, described elsewhere as ${\cal O}(\at \as +
\at^2)$ for brevity.  In particular, contributions of that order are
present in the two--loop corrected EPA masses $(\ov{m}_h^2,\mHbar^2)$.
Also $(\delta^{(1)} \mh^2, \delta^{(1)} \mH^2)$ give ${\cal O}(\at^2
m_t^2)$ terms, induced by the one--loop corrected $(\ov{m}_h^2, \mHbar^2,
\ov{\alpha})$. The question is whether $\ov{m}_h^2 + \delta^{(1)} \mh^2$ 
and $\mHbar^2 + \delta^{(1)} \mH^2$ exhaust the full ${\cal O}(\at \as
m_t^2)$ and ${\cal O}(\at^2 m_t^2)$ corrections to the Higgs masses or
$(\delta^{(2)} \mh^2, \delta^{(2)} \mH^2)$ give additional
contributions of the same order. When $\ma^2 = {\cal O}(\mz^2)$, the
contributions from $(\delta^{(2)} \mh^2,
\delta^{(2)} \mH^2)$ are at most ${\cal O}(\at \as \mz^2)$ 
or ${\cal O}(\at^2 \mz^2)$, i.e. they are suppressed by
${\cal O}(\mz^2/m_t^2)$ with respect to the dominant terms. 
In the case of  $\delta^{(2)} \mh^2$, this
is true also when $\ma^2 > \mz^2$. Indeed, the terms growing with $\ma^2$
in $\delta\Pi_{AA}^{(2)}(\ma^2)$ or $\Delta\Pi_{AA}^{(1)}(\ma^2)$
are actually suppressed by the associated coefficients,
since $\cos^2(\beta -\ov{\alpha})= {\cal O}(\mz^4/\ma^4)$ and
$\sin 2 (\beta -\ov{\alpha})={\cal O}(\mz^2/\ma^2)$.
Therefore the ${\cal O}(\at \as m_t^2)$ and ${\cal O}(\at^2 m_t^2)$ 
corrections to $\mh^2$ are fully accounted for by 
$\ov{m}_h^2 + \delta^{(1)} \mh^2$: no further two--loop 
calculations are needed, as first emphasized in \cite{hh}.
On the other hand, ${\cal O}(\at \as m_t^2)$ and 
${\cal O}(\at^2 m_t^2)$ contributions do generically appear
in $\delta^{(2)} \mH^2$ when  $\ma^2 > \mz^2$. 
In the window  $\mz^2< \ma^2 < m_t^2$, however, such effects
are suppressed at least by a factor ${\cal O}(\ma^2/m_t^2)$.
%
%%%%%%%%%%%%%%%%%%%%%%%%%%%%%%%%%%%%%%%%%%%%%%%%%%%%%%%%%%%%%%%%%%
%
\section{${\cal O}(\at^2)$ two--loop corrections to the 
neutral Higgs boson masses}
\label{section:3}

We shall now describe our two--loop computation of the matrix $(
\Delta{\cal M}^2_S)^{\rm eff}$ at ${\cal O} (\at^2)$, for arbitrary
values of the relevant MSSM input parameters.  The computation is
consistently performed by setting to zero all the gauge couplings (the
so--called gaugeless limit), and by keeping $h_t = \sqrt{4 \pi \at}$
as the only non--vanishing Yukawa coupling. In this limit, the
tree--level (field--dependent) spectrum of the MSSM simplifies
considerably: gauginos and Higgsinos do not mix; charged and neutral
Higgsinos combine into Dirac spinors with degenerate mass eigenvalues,
$|\mu|^2$, where $\mu$ is the Higgs mass parameter in the
superpotential; the only massive SM fermion is the top quark; all
other fermions and gauge bosons have vanishing masses; besides the
stop squarks, to be discussed in detail below, the only sfermion with
non--vanishing couplings is $\tilde{b}_L$, a mass eigenstate with
eigenvalue $m_{\widetilde{b}_L}^2 = m_Q^2$; the lighter CP--even Higgs
boson, $h$, is massless, as can be seen immediately from (\ref{mpma})
with $\ov{m}_\smallz^2=0$; the same is true for the Goldstone bosons;
all the remaining Higgs states, $(H,A, H^\pm)$, have degenerate mass
eigenvalues $\ma^2$. The tree--level mixing angle in the CP--even
sector is just $\alpha=\beta-\pi/2$.

According to Eq.~(\ref{Dms}), $\left( \Delta{\cal M}^2_S
\right)^{\rm eff}$ can be computed by taking the derivatives 
of $V$ with respect to the CP--even and CP--odd fields, 
evaluated at the minimum of $V_{\rm eff}$. In this computation 
we follow closely the strategy developed in \cite{DSZ}, and 
express the field--dependent masses and interaction vertices
that contribute to $V$, and are relevant for our 
calculation, in terms of five field--dependent quantities, 
which can be chosen as follows. The first three quantities 
are the squared masses of the top quark and squarks,
\be
 m_t^2 = h_t^2 |H_2^0|^2 \, ,
\ee
\be
m^2_{\tilde{t}_{1,2}} = \frac{1}{2} \left[ ( m_L^2 + m_R^2) 
\pm  \sqrt{ (m_L^2-m_R^2)^2 + 4 \, |\mix|^2} \right] \, .
\label{mstop}
\ee
In the above equation,
\be
m_L^2 = m_Q^2 + h_t^2 \, |H_2^0|^2 \, ,
\hspace{1cm}
m_R^2 = m_U^2 + h_t^2 \, |H_2^0|^2 \, ,
\ee
\be
\label{mix}
\mix  \; \equiv |\mix|\, e^{i\, \widetilde{\varphi}} \;  =  \;  
h_t\,\left( A_t H_2^0 + \mu H_1^{0\,*}\right) \, ,
\hspace{1.cm}
(0 \leq \widetilde{\varphi} < 2 \pi) \, ,
\ee
where $m_Q^2$, $m_U^2$ and $A_t$ are the soft supersymmetry--breaking
mass parameters of the stop sector. We assume here $\mu$ and $A_t$ to
be real~\footnote{Notice that Eq.~(\ref{mix}) implicitly defines our
conventions on the sign of the parameter $\mu$, and fixes the sign of
$\mu$ in the chargino and neutralino mass matrices. Notice also that
our conventions differ by a sign from those of \cite{hollik}.}, so
that $\langle |\mix| \rangle = (h_t v_2/\sqrt{2}) \, |A_t + \mu
\cot \beta|$ and $\langle e^{i\, \widetilde{\varphi}} \rangle = sign \, 
(A_t + \mu \cot\beta)$, but we do not make any assumption on
the sign of $\mu$ and $A_t$. A fourth quantity is the mixing 
angle $\ths$ in the stop sector ($0 \leq \ths < \pi/2$), where:
\be
\sin 2 \ths = \frac{2 \, |\mix|}{\msqu - \msqd} \, ,
\;\;\;\;\;
\cos 2 \ths = \frac{m_L^2-m_R^2}{\msqu - \msqd} \, .
\label{sint}
\ee
Notice that the usual field--independent definition for the angle
$\thz$ that diagonalizes the stop mass matrix at the minimum,
\be
\label{sinmin}
\sin 2 \thz = \frac{2\,m_t\,(A_t + \mu\cot\beta)}{\diff} \, ,
\;\;\;\;\;
\cos 2 \thz = \frac{m_Q^2 - m_U^2}{\diff} \, ,
\ee
leads instead to $-\pi/2 \leq \thz < \pi/2$. A fifth quantity can be
chosen to be a function of the phase difference $\varphi -
\widetilde{\varphi}$, where $\varphi$ is the phase in the complex top
mass, $\mixt \; = \; h_t\, H^0_2 \; \equiv \; |\mixt|
\, e^{i\,\varphi}$, with $0 \leq \varphi < 2 \pi$ such that $\langle 
|\mixt| \rangle = h_t v_2 / \sqrt{2}$ and $\langle \varphi \rangle = 0$.
A convenient choice is $\cpp \equiv \cos\,(\varphi - \tilde{\varphi})$, 
with $\langle \cpp \rangle = \pm 1$. 

The classes of Feynman diagrams that contribute to the two--loop
expression of $V$ and affect the ${\cal O}(\at^2)$ calculation 
of the neutral Higgs boson masses~\footnote{An explicit expression 
of the ${\cal O}(\at^2)$ corrections to the effective potential can 
be found in Eq.~(D.6) of \cite{ez2}. To extend its validity to the 
general case discussed in the present paper, such equation
should be modified  in the following way: take the couplings 
$\lambda^2_{H^0_n \tilde{t}_i \tilde{t}_j}$ and 
$\lambda^2_{H^+_n \tilde{t}_i \tilde{b}_L}$, appearing in 
the last line of Eq.~(D.6); in the corresponding definitions,
Eq.~(B.6), multiply the terms $\mt \, \s2t X_t$ and $\mt \,
\s2t Y_t$ by the factor $\cpp$.} are shown in Fig.~1. 
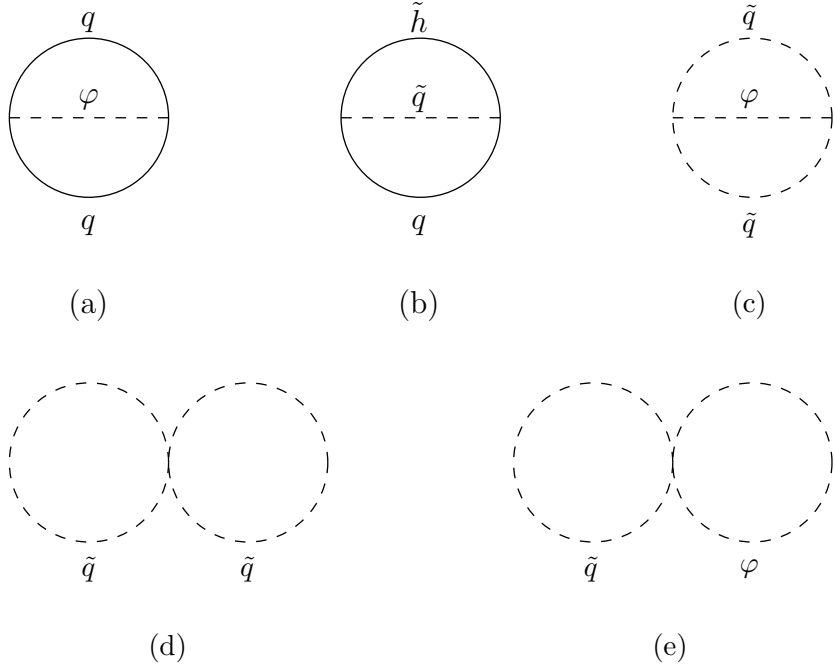
\begin{figure}[t]
\begin{center}
\begin{picture}(330,280)(0,0)
%
%\Text(0,75)[]{.}
%\Text(330,75)[]{.}
%\Text(0,205)[]{.}
%\Text(330,205)[]{.}
%
\CArc(40,205)(30,0,360)
\DashLine(10,205)(70,205){4}
\Text(40,165)[]{\large $q$}
\Text(40,242)[]{\large $q$}
\Text(40,213)[]{\large $\varphi$}
\Text(40,135)[]{\large (a)}
\CArc(165,205)(30,0,360)
\DashLine(135,205)(195,205){4}
\Text(165,165)[]{\large $q$}
\Text(165,244)[]{\large $\tilde{h}$}
\Text(165,213)[]{\large $\tilde{q}$}
\Text(165,135)[]{\large (b)}
\DashCArc(290,205)(30,0,360){4}
\DashLine(260,205)(320,205){4}
\Text(290,165)[]{$\tilde{q}$}
\Text(290,244)[]{$\tilde{q}$}
\Text(290,213)[]{$\varphi$}
\Text(290,135)[]{(c)}
\DashCArc(40,75)(30,0,360){4}
\DashCArc(100,75)(30,0,360){4}
\Text(40,35)[]{$\tilde{q}$}
\Text(100,35)[]{$\tilde{q}$}
\Text(70,5)[]{(d)}
\DashCArc(230,75)(30,0,360){4}
\DashCArc(290,75)(30,0,360){4}
\Text(230,35)[]{$\tilde{q}$}
\Text(290,35)[]{$\varphi$}
\Text(260,5)[]{(e)}
\end{picture}
\end{center}
\vspace{-0.5cm}
\caption{ The classes of Feynman diagrams that 
contribute to the two--loop effective potential and
affect the ${\cal O}(\alpha_t^2)$ calculation of the
neutral Higgs boson masses [$q=(t,b)$, $\varphi 
= (H, h, G, A, H^\pm, G^\pm)$, $\tilde{h} = 
(\tilde{h}^0_{1,2}, \tilde{h}^\pm)$, $\tilde{q} = 
(\tilde{t}_1, \tilde{t}_2, \tilde{b}_L)$].} 
\label{fig1}
\end{figure}

Introducing a wave function renormalization (w.f.r.) matrix $Z$ for the
Higgs fields, we can write 
\be
\label{dmsren}
\left(\Delta {\cal M}^2_S\right)^{\rm eff} =  
\sqrt{Z} \, \left( \Delta {\cal \widehat M}^2_S 
\right)^{\rm eff}  \sqrt{Z} \, ,
\ee
where
\bea
\label{dms11}
\hspace{-1cm}\left(\Delta {\cal \widehat M}^2_S\right)_{11}^{\rm eff} 
& = &
\frac{1}{2} \, h_t^2 \,\mu^2 \,\s2t^2  \,
\left(F_3 + 2 \, \Delta_\mu F_3 \right)  \, , \\
\label{dms12}
\hspace{-1cm}\left(\Delta {\cal \widehat M}^2_S\right)_{12}^{\rm eff} 
& = &
h_t^2 \,\mu\, m_t\, \s2t \,  \left( F_2 +  \Delta_\mu F_2 \right)
+ \frac{1}{2}\, h_t^2\, A_t \,\mu \, \s2t^2 \, \left(F_3 
+ \Delta_{A_t} F_3 + \Delta_\mu F_3 \right) \, , \\
\label{dms22}
\hspace{-1cm}\left(\Delta {\cal \widehat M}^2_S\right)_{22}^{\rm eff} 
& = &
2\, h_t^2\, m_t^2\, F_1
+ 2\, h_t^2\, A_t\, m_t\, \s2t\, 
\left( F_2 + \Delta_{A_t} F_2 \right)
+ \frac{1}{2}\, h_t^2\, A_t^2\, \s2t^2\, 
\left(F_3 + 2\, \Delta_{A_t} F_3 \right) \, .
\eea
In Eqs.~(\ref{dms11})--(\ref{dms22}), $h_t^2 = 2 \mt^2/(v^2\, \sb^2)$
is the top Yukawa coupling, and $\s2t \equiv \sin 2\thz$ refers to the
field--independent stop mixing angle defined in Eq.~(\ref{sinmin}).
The functions $F_i$ ($i=1,2,3$) are written as $F_i = \widetilde{F}_i
+ \Delta \widetilde{F}_i$, where $\widetilde{F}_i$ are combinations of
derivatives of the effective potential with respect to the
field--dependent quantities, whose explicit expressions are given in
Eqs.~(28)--(30) of \cite{DSZ}. The terms $\Delta \widetilde{F}_i$
include the renormalization of the common factors multiplying $F_i$
(i.e. $h_t^2,\,\mt,\,\s2t$), as well as the renormalization of the
mass parameters appearing in the one--loop parts of the $F_i$, while $
\Delta_{A_t} F_i$ and $\Delta_\mu F_i$ encode the renormalization of
the factors that are not common~\footnote{Indeed, the terms $F_i$
in Eqs.~(\ref{dms11})--(\ref{dms22}) are multiplied by different
combinations of $\mu$ and $A_t$. These two parameters have different
${\cal O}(\at)$ renormalizations that cannot be absorbed into the
$F_i$, and are then separately taken into account. In Ref.~\cite{DSZ},
which deals with the ${\cal O} (\at \as)$ corrections, the expressions
analogous to Eqs.~(\ref{dmsren})--(\ref{dms22}) contain neither the
w.f.r.  $Z$, nor the terms $\Delta_\mu F_i$, since in such case the
w.f.r.~is not needed and the parameter $\mu$ does not receive ${\cal
O}(\as)$ corrections. In \cite{DSZ}, what we call here $(\Delta {\cal
\widehat M}^2_S)^{\rm eff}$ and $ \Delta_{A_t} F_i$ are indicated as
$(\Delta {\cal M}^2_S)$ and $\Delta F_i$, respectively.}.

We performed the evaluation of ($F_1, \, F_2,\,F_3$) following the
strategy of \cite{DSZ}, namely computing directly the derivatives of
$V$, without evaluating $V$ explicitly. Before discussing the ${\cal
O}(\at^2)$ two--loop terms we recall first, for completeness, the
one--loop ${\cal O}(\at)$ result \cite{first}:
\be
\label{fi1l}
F_1^{\rm 1\ell} = \frac{N_c}{16\,\pi^2}\ln\frac{\msqu \msqd}{m_t^4}\,,
\hspace{0.7cm}
F_2^{\rm 1\ell} = \frac{N_c}{16\,\pi^2}\ln\frac{\msqu}{ \msqd}\,,
\hspace{0.7cm}
F_3^{\rm 1\ell} = \frac{N_c}{16\,\pi^2} \left( 2 -
\frac{\msqu+\msqd}{\msqu-\msqd}
\,\ln\frac{\msqu}{ \msqd} \right) \, ,
\ee
where $N_c = 3$ is a color factor.

We organize the presentation of our two--loop results as follows. In
this section we give the results assuming that all the parameters
(masses, VEVs, mixing angles and Yukawa couplings) appearing in the
one--loop term of Eqs.~(\ref{dms11})-(\ref{fi1l}) are expressed in the
$\ov{\rm DR}$ scheme. In the next section we give the expressions for
the shifts of these parameters that allow to translate the $\ov{\rm
DR}$ result into other renormalization schemes. In particular, we
specialize the shifts to a convenient renormalization scheme, and we
comment on alternative choices in the existing literature.

Since the general analytic expressions of $(F_1,\,F_2,\,F_3)$ in
Eqs.~(\ref{dms11})--(\ref{dms22}) are very long, we choose not to
display them in print. Instead, they can be obtained upon request as a
Mathematica or Fortran code. Here we present explicit analytical
formulae for $m_Q = m_U \equiv M_S$, and neglecting ${\cal O}
(\mt/M_S)$ corrections, but without making any assumption on the size
of $\ma$. In this case, the stop masses are given, in the gaugeless
limit, by $m_{\tilde{t}_{1,2}}^2 = M_S^2 + \mt^2 \pm \mt |X_t|$, where
$X_t = A_t + \mu\cot\beta$, and $\s2t^2=1$. Setting
consistently the bottom mass to zero, the relevant sbottom mass is
$m_{\tilde{b}_L}^2 = M_S^2$.  We denote by $(\hat{F}_{1}^{\rm
2\ell},\hat{F}_{2}^{\rm 2\ell}, \hat{F}_{3}^{\rm 2\ell})$ the
expressions for the two--loop ${\cal O}(\at^2)$ part of the $F_i$
functions obtained assuming that the one--loop part is evaluated in
terms of $\ov{\rm DR}$ quantities.  In units of $h_t^2
\,N_c/(16\,\pi^2)^2$, we have:

\bea
\label{f12l}
\hat{F}_{1}^{\rm 2\ell} & = &
-\frac{6}{\epsilon}\left(\ln \frac{\mt^2}{M_S^2}
+\frac{X_t\,A_t}{M_S^2}\right)
-6\logT \ln \frac{\mt^2}{M_S^2}
- \sb^2\,{X_t^2 \over M_S^2}\,\left(6-15\,\logT\right) 
- \sb^2\,\frac{\pi^2}{3} \nn\\
\hspace{-3cm} &&\nn\\
&+&
6\,\frac{\mhg^2}{M_S^2}\,\left(1-\logT\right)
- 2\,\left(\frac{8\,\mhg^2 - 5\,M_S^2}{\mhg^2 - M_S^2}- 
\frac{3\,\mhg^4}{\left( \mhg^2 - M_S^2 \right)^2} 
\ln \frac{\mhg^2}{M_S^2} \right)\,\logTt
\nn \\
&-&    \frac{2\,\mhg^2\,
\left( 3\,\mhg^4 + \,\mhg^2\,M_S^2 +2 M_S^4 \right) }
{M_S^2 \,\left( \mhg^2 - M_S^2 \right)^2 }\ln \frac{\mhg^2}{M_S^2}
-\frac{4\,\left( 2\,\mhg^4 + 2\,\mhg^2\,M_S^2 - M_S^4 \right) 
       }{ \left( \mhg^2 - M_S^2 \right)^2 } 
\, \li2 \left(1 - \frac{\mhg^2}{M_S^2}\right) \nn\\
\hspace{-3cm} &&\nn\\
&+&\cb^2\,\left\{
\frac{4\,\A0}{\A0-4\,\t}\logAt
- 2\,\li2\left(1-\frac{\,\A0}{\t}\right)
+ 2\,\frac{(\A0-6\,\t)}{(\A0-4\,\t)}\,
\phi\left({\ma^2 \over 4\,\mt^2} \right)
\right.\nn\\
\hspace{-3cm} &&\nn\\
\hspace{-3cm} &&\hspace{1cm}
-3\,\frac{\,\A0}{\T}\left(1-\logA0\right)
-5\,\phi \left({\ma^2 \over 4 M_S^2 } \right)
\,  +\frac{6\, X_t\, Y_t}{M_S^2}\, \,\logT \nn\\
&&\nn\\
&&+ \left.
\frac{2\, X_t\, Y_t + Y_t^2}{M_S^2} \left[3\,\logA0-3
+\frac{10\,\T}{\A0-4\,\T}\,\logAT
-2\,\frac{M_S^2 \left(\A0+\T\right)}{\ma^2 \left(\A0-4\,\T\right)}
\phi \left({\ma^2 \over 4 \, M_S^2 } \right)
\,\right] \right\} \, , \nn\\
\eea
%%%%%%%%%%%%%%%%%%%%%%%%%%%%%%%%%%%%%
 \bea
\label{f22l}
\frac{M_S}{m_t}\,\s2t\,\hat{F}_{2}^{\rm 2\ell}& = &
\frac{1}{\epsilon}\,\frac{X_t^2\,A_t}{M_S^3}
+ 5\,\sb^2\,{X_t^3 \over M^3_S}\,\left(1 - \,\logT\right) \nn \\
\hspace{-3cm} &&\nn\\
\hspace{-3cm} &+& {2\,X_t \over M_S^3}\,
\left[ 3 \,(4\, M_S^2 + \mhg^2) \ln\frac{M_S^2}{Q^2}
+ \frac{\mhg^4\, \left( 3\,\mhg^2 - 5\,M_S^2 \right)}
{\left( \mhg^2 - M_S^2 \right)^2 } \ln\frac{\mhg^2}{M_S^2} 
- \frac{\left( 3\,\mhg^4 - \mhg^2\,M_S^2 - 4\,M_S^4 \right) }
{\left( \mhg^2 - M_S^2 \right) } \right]\nn\\
\hspace{-3cm} &&\nn\\
&+& \cb^2\,\left\{
{X_t \over M_S}\,\left[ 3\,\frac{\A0+\T}{\T} 
- \frac{3\,\A0-5\,\T}{\T}\,\logA0 
\right.\right.\nn\\
\hspace{-3cm} &&\nn\\
\hspace{-3cm} &&\left.\hspace{2cm}
- 11\,\logT+\frac{18\,\T}{\A0-4\,\T}\,\logAT
- 2\,\frac{M_S^2 \left(2\,\A0+\T\right)}{\ma^2\left(\A0-4\,\T\right)}\,
\phi \left({\ma^2 \over 4 \, M_S^2 } \right)
\right]\nn\\
\hspace{-3cm} &&\nn\\
\hspace{-3cm} &&
+{Y_t \over M_S}\,\left[6\,\logA0 - 6 + \frac{20\,\T}{\A0-4\,\T}\,\logAT
-4\,\frac{M_S^2 \left(\A0+\T\right)}{\ma^2 \left(\A0-4\,\T\right)}\,
\phi \left({\ma^2 \over 4 \, M_S^2 } \right)
\right]\nn\\
\hspace{-3cm} &&\nn\\
\hspace{-3cm} &&
+ {X_t^2 \, Y_t \over M_S^3}\,\left[2 -\logA0-\logT
- \frac{3\,\T}{\A0-4\,\T}\right.\nn\\
\hspace{-3cm} &&\nn\\
\hspace{-3cm} && \hspace{2cm}\left.
+ \frac{\T\,(\A0+14\,\T)}{(\A0-4\,\T)^2}\,\logAT
- 2\,\frac{M_S^4\,(2\,\A0+\T)}{\ma^2\,(\A0-4\,\T)^2}
\,\phi \left({\ma^2 \over 4 \, M_S^2 } \right)
\right]\nn\\
\hspace{-3cm} &&\nn\\
\hspace{-3cm} &&
+{X_t \, Y_t^2\over M_S^3}\,\left[3\left(1-\logA0\right)-
\frac{5\,\T}{\A0-4\,\T}
-\frac{\T\,(\A0-34\,\T)}{(\A0-4\,\T)^2}\,\logAT\right.\nn\\
\hspace{-3cm} &&\nn\\
\hspace{-3cm} &&\hspace{2cm}\left.\left.
-\frac{2\,M_S^4\,(2\,\A0+7\,\T)}{\ma^2(\A0-4\,\T)^2}\,
\phi \left({\ma^2 \over 4 \, M_S^2 } \right)
\right]\,\right\} \, , 
\eea
%%%%%%%%%%%%%%%%%%%%%%%%%%%%%%%%%%
\bea
\label{f32l}
\frac{\T}{\t}\,\hat{F}_{3}^{\rm 2\ell} & = &
\frac{2}{\epsilon}\,\frac{X_t^2}{M_S^2} + 
{X_t^2 \over M_S^4}\,\left[-(14 \,M_S^2+ 4\,\mhg^2) \ln\frac{M_S^2}{Q^2} 
   - \frac{2\,\mhg^4\, \left( 2\,\mhg^2 - 3\,M_S^2 \right)}
 {\left( \mhg^2 - M_S^2 \right)^2 }\ln\frac{\mhg^2}{M_S^2} 
 \right. \nn\\
\hspace{-3cm} &&\nn\\
\hspace{-3cm}&& + \left.
\frac{ 4\,\mhg^4 +5 \,\mhg^2\,M_S^2 - 11\,M_S^4}
{\left( \mhg^2 - M_S^2 \right)}\right] 
-2\,\sb^2\,{X_t^4 \over M_S^4}\,\left(1-\logT\right) \nn \\
&+& \cb^2\,\left\{
{X_t^2 \over M_S^2}\,\left[4\,\logT 
-\frac{2\,\A0}{\T}\,\left(1-\logA0\right)-6\right]
\right.\nn\\
&&\nn\\
&& + {X_t \,Y_t\over M_S^2} \,\left[4\,\left(1-\logA0\right)
-\frac{12\,\T}{\A0-4\,\T}
+\frac{4\,\T\,(\A0+14\,\T)}{(\A0-4\,\T)^2}\,\logAT \right. \nn \\
&& \left.
-\frac{8\,M_S^4\,(2\,\A0+\T)}{\ma^2(\A0-4\,\T)^2}\,
\phi \left({\ma^2 \over 4 \, M_S^2 } \right)\,\right]
-{2\,X_t^2 \,Y_t^2 \over M_S^4}\,\left[
\left(1-\logA0\right)
+\frac{\T}{\A0}\left(\frac12-\logAT\right) \right.
\nn\\
\hspace{-3cm} &&\nn\\
\hspace{-3cm} &&
+\frac{5\,\Tq\,(\A0+2\T)}{\A0\,(\A0-4\,\T)^2}
-\frac{2\,\Tq\,(2\,\Aq+11\,\A0\,\T+14\,\Tq)}
{\A0\,(\A0-4\,\T)^3}\logAT  \nn\\
\hspace{-3cm} &&\nn\\
\hspace{-3cm} &&\left.\left.
+ \frac{12\,M_S^6\,(\Aq-\A0\,\T+3\,\Tq)}{\ma^4\,(\A0-4\,\T)^3}
\phi \left({\ma^2 \over 4 \, M_S^2 } \right)
\,\right]\,\right\} \, ,
\eea
where $\epsilon= (4-n)/2$ and $n$ is the dimension of the
space--time. We notice that all the double poles $1/\epsilon^2$ 
generated by two--loop diagrams have cancelled in the functions 
$\hat{F}_{i}^{\rm2\ell}\, (i=1,2,3)$, whereas a few $1/\epsilon$ poles 
have survived. Also, $Q$ is the $\ov{\rm DR}$ renormalization
scale, $Y_t = A_t - \mu \tan \beta$, ${\rm{Li_2}} (z) = - \int_0^z dt
\, [\ln (1-t)] / t$ is the dilogarithm function, and
\be
\phi(z) = \left\{
       \begin{tabular}{ll}
       $4 \sqrt{{z \over 1-z}} ~{\rm Cl}_2 ( 2 \arcsin \sqrt z )$ \, , &  
       $(0 < z < 1)$ \, , \\ \\
       ${ 1 \over \lambda} \left[ - 4 {\rm Li_2} ({1-\lambda \over 2}) +
       2 \ln^2 ({1-\lambda \over 2}) - \ln^2 (4z) +\pi^2/3 \right] $ \, ,
       & $(z \ge 1)$ \,,
       \end{tabular}
       \label{e2.15c}
\right. 
\ee 
where ${\rm Cl}_2(z)= {\rm Im} \,{\rm Li_2} (e^{iz})$ is the Clausen 
function and $\lambda = \sqrt{1 - (1 / z)}$.

Under the same approximations of Eqs.~(\ref{f12l})--(\ref{f32l}), and
in the same units, the ${\cal O}(\at^2)$ contributions to $\Delta_\mu
F_i$ and $\Delta_{A_t} F_i$ are:
\be
\label{DmF2}
\s2t\, \Delta_{A_t}\hat{F}_2 \;\; = \;\; 4\,\s2t\, \Delta_\mu \hat{F}_2\;\;  
=  \;\; 
\frac{12\,m_t\,X_t}{M_S^2} \, \left( \frac{1}{\epsilon} -\logT\right)\,, 
\ee

\be
\label{DAF2}
\Delta_{A_t} \hat{F}_3 \;\; = \;\; 4\,\Delta_\mu \hat{F}_3 \;\; = \;\;
- \frac{4\,\t\,X_t^2}{M_S^4}\, \left(\frac{1}{\epsilon} + 1-\logT\right)\,.
\ee
Finally, the w.f.r. matrix $Z$ induces additional contributions to the
entries of $\left(\Delta {\cal M}^2_S\right)^{\rm eff}$. 
In the $\ov{\rm DR}$ scheme $Z$ is diagonal, $(Z)_{ij} = Z_i\, \delta_{ij}$,
and only $Z_2$ is different from 1, since we are neglecting the gauge 
couplings and the bottom Yukawa coupling. The entries (1,2) and (2,2) 
get then additional corrections that in units of
$h_t^4\,N_c^2/(16\,\pi^2)^2$, and in the same approximations of
Eqs.~(\ref{f12l})--(\ref{DAF2}), read:
\bea
\label{DZ12}
\sqrt{Z_2} \,\left(\Delta {\cal \widehat M}^2_S\right)_{12}^{\rm eff} 
&=& 
\;\;\:\t\left[
- \frac{\mu\,X_t}{\T}\,\left(\frac{1}{\epsilon} - \logT\right) 
+ \frac{A_t\,\mu\,X_t^2}{6\,\Tq}\,
\left(\frac{1}{\epsilon}+1-\logT\right)\right] \, ,\\
&&\nn\\
Z_2 \,\left(\Delta {\cal \widehat M}^2_S\right)_{22}^{\rm eff} 
&=& 
2\,\t\,\left[
\left( \frac{2}{\epsilon} - \logT - \logt\right)
\ln\frac{m_t^2}{M_S^2}
- \frac{2\,A_t\,X_t}{\T}\,\left(\frac{1}{\epsilon} - \logT\right) 
\right.\nn\\
& & \left.\hspace{1.5cm}
+ \frac{A_t^2\,X_t^2}{6\,\Tq}\,
\left(\frac{1}{\epsilon} + 1-\logT\right)\right] \, .
\label{DZ22} 
\eea

Some comments on Eqs.~(\ref{f12l})--(\ref{DZ22}) are in order.  It can
be checked that, after the inclusion of the w.f.r.
[Eqs.~(\ref{DZ12})--(\ref{DZ22})], all the leftover divergent contributions 
to $\left(\Delta {\cal M}^2_S\right)^{\rm eff}$ do indeed cancel out. 
We have verified that an analogous cancellation takes place also in 
the general case.
The matrix $\left(\Delta {\cal M}^2_S \right)^{\rm eff}$ shows an explicit
dependence on the renormalization scale $Q$, due not only to our
choice of expressing the various parameters in the $\ov{\rm DR}$
scheme, but also to the fact that the entries of $\Gamma_S(p^2)$ are
not physical quantities. Only the solutions to Eq.~(\ref{3.49a})
should be $Q$--independent, once the quantities they depend upon are
expressed in terms of physical observables. Then, to obtain a
$Q$--independent result for the eigenvalues, we should also take into
account, in the perturbative diagonalization of $\Gamma_S (p^2)$, the
contribution induced by $\left(\Delta {\cal M}^2_S\right)^{p^2}$ in
Eq.~(\ref{mbar}). 

We have also checked that, for $\ma = M_S \gg \mz$ and neglecting ${\cal O}
(m_t/M_S)$ corrections, the ${\cal O}(\at^2)$ contribution to $\mh^2$
due to $\left(\Delta {\cal M}^2_S \right)^{\rm eff}$, namely $\cb^2
\left(\Delta {\cal M}^2_S \right)^{\rm eff}_{11} + 2 \sb\cb \left(
\Delta {\cal M}^2_S \right)^{\rm eff}_{12} + \sb^2 \left(\Delta 
{\cal M}^2_S \right)^{\rm eff}_{22}$, given by 
Eqs.~(\ref{f12l})--(\ref{DZ22}), reduces to Eq.~(17) of Ref.~\cite{ez2}.
%
%%%%%%%%%%%%%%%%%%%%%%%%%%%%%%%%%%%%%%%%%%%%%%%%%%%%%%%%%%%%%%%%%%
%
\section{Input parameters and renormalization schemes}
\label{section:3bis}

The general results of our calculation can be combined with the
complete one--loop calculation and with the calculation of the ${\cal
O}(\at \as)$ two--loop corrections, to give the most accurate
determination of the neutral CP--even Higgs masses available to
date. Here and hereafter, when referring to the masses of the CP--even
Higgs bosons, $(\mh,\mH)$, we will always mean their physical
($\equiv$pole) masses. Before presenting our numerical results, we
must discuss the input parameters that need to be defined beyond the
tree level. They are the parameters ($\ma,\mz,\tan \beta$) that appear
already in the tree--level result, and the parameters of the top and
stop sectors that appear first in the ${\cal O} (\at)$ one--loop
correction. Other quantities, such as the gluino mass $m_{\tilde{g}}$,
or the strong coupling constant $\as$, appear only at the two--loop
level and do not require further specification. The same is true for
the parameters, such as the $SU(2)_L \times U(1)_Y$ gaugino masses
$(M_2,M_1)$ or the remaining sfermion masses, that appear only in the
complete one--loop correction but do not appear in the ${\cal O} (\at
\as + \at^2)$ two--loop correction.

According to the discussion of Section \ref{section:2}, the parameters 
$\ma$ and $\mz$ are chosen to be the physical masses, while the result 
for $( \Delta{\cal M}^2_S)^{\rm eff}$ presented in Section \ref{section:3} 
is  expressed in terms of $\ov{\rm DR}$ quantities. We give now the 
formulae that allow to express our results in terms of input parameters 
given in a different renormalization scheme, which we indicate 
generically as $R$. In the case of the ${\cal O}(\at \as)$ corrections, 
the procedure was explained in \cite{DSZ}, but we shall complete the
discussion here to take into full account the ${\cal O} (\at \as + 
\at^2)$ corrections.

To obtain the ${\cal O}(\at \as + \at^2)$ correction in the
$R$ scheme, we have just to shift the parameters appearing
in the one--loop term. Indicating, generically,  a quantity in the 
$\ov{\rm DR}$ scheme as $x^{\ov{\rm DR}}$, and the same quantity in 
the $R$ scheme as $x$, we can write the one--loop relation $x = 
x^{\ov{\rm DR}} - \delta x$. Then, once the one--loop contribution 
is evaluated in terms of $R$ quantities, the two--loop ${\cal O}(\at 
\as + \at^2 )$ corrections in the $R$ scheme  can be obtained through:
\bea
F_1^{\rm 2\ell} & = & \hat{F}_1^{\rm 2\ell} 
+\frac{N_c}{16\,\pi^2}\,
\left(\frac{\delta\msqu}{\msqu} + \frac{\delta\msqd}{\msqd} 
- 4\, \frac{\delta m_t}{m_t} \right)
+ 2\, \left( 2\, \frac{\delta m_t}{m_t} -  \frac{\delta v}{v}
- \frac{\delta \sb}{\sb} \right) \, F_1^{\rm 1\ell} \, ,
\label{f12lr}\\
F_2^{\rm 2\ell} & = & \hat{F}_2^{\rm 2\ell} 
+\frac{N_c}{16\,\pi^2}\,
   \left(\frac{\delta\msqu}{\msqu} - \frac{\delta\msqd}{\msqd}\right) 
+ \left( 3\, \frac{\delta m_t}{m_t} + \frac{\delta \s2t}{\s2t} 
- 2\, \frac{\delta v}{v} - 2\, \frac{\delta \sb}{\sb} \right)
\, F_2^{\rm 1\ell} \, ,
\label{f22lr}\\
F_3^{\rm 2\ell} & = & \hat{F}_3^{\rm 2\ell}
+ \frac{N_c}{16\,\pi^2}\,
\left(2 \,\frac{\msqu \msqd}{(\msqu-\msqd)^2}\ln\frac{\msqu}{\msqd}
- \frac{\msqu+\msqd}{\msqu-\msqd}\right)\;
\left(\frac{\delta\msqu}{\msqu} - \frac{\delta\msqd}{\msqd} \right)\nn \\
&& \hspace{0.7cm} + 2\, \left(\frac{\delta m_t}{m_t} 
+ \frac{\delta \s2t}{\s2t} -  \frac{\delta v}{v}
-  \frac{\delta \sb}{\sb} \right)\, F_3^{\rm 1\ell}
\label{f32lr} \, , \\
&&\nn
\eea
%%%%%%%%%%%%%%%%%%%%%%%%%%%%%%%%%
\be
\Delta_{A_t} F_i =  \Delta_{A_t} \hat{F}_i + 
\frac{\delta A_t}{A_t} \, F_i^{\rm 1\ell} \, ,
\;\;\;\;
\Delta_{\mu} F_i =  \Delta_{\mu} \hat{F}_i + 
\frac{\delta\, \mu}{\mu} \, F_i^{\rm 1\ell} \, ,
\;\;\;\;
(i=2,3) \, ,
\label{dfam23}
\ee
where all the quantities that appear in Eqs.~(\ref{f12lr})--(\ref{dfam23}) 
are meant in the $R$ scheme. We notice that the shifts appearing in 
Eqs.~(\ref{f12lr})--(\ref{dfam23}) are not all independent, because of 
the relation (\ref{sinmin}).

We choose to express our result identifying $\mt$ with the pole
top mass, while the top Yukawa coupling will be defined indirectly by
$h_t = \sqrt{2}\,m_t / (v \sb)$. For the electroweak
symmetry--breaking parameter $v$, we will use the value obtained in
terms of the precisely known muon decay constant $G_{\mu}$, according
to the relation $v = (\sq2\, G_{\mu})^{-1/2} = 246.218$ GeV. These
choices set $\delta \mt$ and $\delta v$ to be
\be
\label{dmtdv}
\delta \mt = {\rm Re}\,\hat{\Sigma}_{t}(m_t)\,,\hspace{1cm}
\frac{\delta v}{v} = 
\frac{1}{2} \,\frac{\hat{\Pi}^{\;T}_{WW}(0)}{\mw^2}\,,
\ee
where $\hat{\Sigma}_{t}(m_t)$ is the finite part of the top 
quark self-energy, evaluated at an external momentum equal 
to the top mass, and $\hat{\Pi}^{\;T}_{WW}(0)$ is the finite
transverse part of the W-boson self-energy, evaluated at zero 
external momentum.

The parameter $\tan \beta$ appears in the tree--level result, in the
stop mass matrix and also, implicitly, in $h_t$. We recall that, at
the tree level and in the gaugeless limit, the stop masses and mixing
angle depend on $\mt$ and five more parameters, $(m_Q, m_U, A_t, \mu,
\tan \beta)$. The last three enter the stop mass matrix only in the
combination $X_t$, so that $(\mtu, \mtd, \s2t)$ are completely defined
in terms of the three quantities $(m_Q, m_U, X_t)$. Yet, the dependence
of $\left( \Delta {\cal M}^2_S \right)^{\rm eff}$ on $(A_t, \mu, \tan
\beta)$ is not through the combination $X_t$ but is more complicated, 
as can be seen from Eqs.~(\ref{dms11})--(\ref{dms22}). In the case of
the ${\cal O} (\at \as)$ corrections, since $\mu$ and $\tan \beta$ do
not receive ${\cal O} (\as)$ one--loop corrections, the specification
of the three parameters of the stop sector, together with $\mt$, is
actually sufficient to fully determine $\left(\Delta {\cal
M}^2_S\right)^{\rm eff}$. The situation is different for the ${\cal O}
(\at^2)$ corrections. Since $A_t$, $\mu$ and $\tan\beta$ are all
corrected by ${\cal O} (\at)$ one--loop effects, their individual
definition is required, in addition to the specification of
$(m_t,v,\mtu,\mtd)$. However, because of the relation (\ref{sinmin}),
we can actually trade one of the parameters $(A_t,\mu,\tan\beta)$ for
$\s2t$.

While there is a well known OS definition for the masses, and an OS
definition for the stop mixing angle can be also conceived, it is not
clear what meaning should be assigned to an OS definition of
$(A_t,\mu,\tan\beta)$. For instance, they could be related to specific
physical amplitudes. However, given our present ignorance of any
supersymmetric effect, such a choice does not seem particularly
useful.  Then, it may seem simpler to assign all the parameters
related with the stop sector, i.e. $(m_Q,m_U,A_t,\mu,\tan\beta)$, in
the $\ov{\rm DR}$ scheme, at a chosen reference scale $Q_0$. However,
as noticed in \cite{DSZ}, in the $\ov{\rm DR}$ scheme the limit of
heavy gluino is very violent, since the ${\cal O} (\at \as)$
corrections to the Higgs masses include terms proportional to $\mgl$
and $\mgl^2$. This powerlike behavior is actually not present if the
stop masses and mixing angle are assigned in a suitable OS scheme,
since it is cancelled by the relevant shifts in
Eqs.~(\ref{f12lr})--(\ref{dfam23}). In fact, with the latter choice
the ${\cal O}(\at \as)$ corrections grow only logarithmically as
$\mgl$ increases. In this situation, we think that the use of OS stop
masses and mixing angle should be preferred, while for the quantities
$\mu$ and $\tan\beta$ we are still going to employ the $\ov{\rm DR}$
definition, at a reference scale that we choose near the present
central value of the top quark mass, $Q_0 = 175$~GeV. In this
framework, we treat $A_t$ as a derived quantity, obtained through
(\ref{sinmin}), whose shift is given by
\be
\delta A_t = \left( 
\frac{\delta\msqu - \delta\msqd}{\msqu - \msqd} +
\frac{\delta \s2t}{\s2t} - \frac{\delta m_t}{m_t}\right) 
\,(A_t + \mu \cot\beta) -\cot\beta \,\delta \mu - \mu \,\delta \cot\beta~~.
\label{dAt}
\ee

According to the above discussion, we are going to take $\delta \mu =
\delta \beta =0$. The requirement of OS stop masses specifies $\delta
\msqu$ and $\delta\msqd$ uniquely as
\be
\delta\msqu = {\rm Re}\,\hat{\Pi}_{11}(\msqu)\,,\hspace{1cm}
\delta\msqd = {\rm Re}\,\hat{\Pi}_{22}(\msqd)\,,
\ee
where $\hat{\Pi}_{ii}(m^2_{\tilde{t}_i})\,$ (i = 1,2) are the finite
parts of the diagonal stop self-energies, in the mass eigenstate
basis, evaluated at external momenta equal to the corresponding
masses. For the stop mixing angle, several OS prescriptions are
present in the literature (for a discussion see, e.g.,
\cite{guasch,hollik,yam,en2} and references therein). We find the
following `symmetrical' definition \cite{guasch,yam}
\be
\delta\thz =  \frac{1}{2}\, 
\frac{\hat{\Pi}_{12}(\msqu) + \hat{\Pi}_{12}(\msqd)}{\msqu-\msqd} \, ,
\label{detheta}
\ee
such that  $\delta \s2t/\s2t =  2 \, \cot 2\thz \, \delta\thz$,
suitable for our ${\cal O}(\at^2)$ calculation, while the choice
\be
\label{thalt}
\delta\thz =
\frac{\hat{\Pi}_{12}(\msqu)}{\msqu-\msqd} \, ,
\ee
employed in \cite{hollik} for the calculation of the ${\cal O} (\at 
\as)$ corrections, 
seems less justified at ${\cal O} (\at^2)$, since the 
divergent parts in the `bare' version of Eq.~(\ref{thalt}) do not match.
In Eqs.~(\ref{detheta})--(\ref{thalt}), $\hat{\Pi}_{12}(p^2)$ stands for 
the off--diagonal stop self--energy.

Explicit formulae for the  shifts relevant to our calculation are 
provided in the Appendix. These formulae include only the contributions 
that come from the top--bottom sector of the MSSM, and are specialized 
to the gaugeless limit we are considering in this work. General 
expressions for the top and stop self--energies at one loop can be 
found in the literature \cite{bmpz,donini}. 
%
%%%%%%%%%%%%%%%%%%%%%%%%%%%%%%%%%%%%%%%%%%%%%%%%%%%%%%%%%%%%%%%%%%
%
\section{Numerical results}
\label{section:4}

In this section we discuss the effect of our ${\cal O}(\at^2)$ two--loop
corrections on the masses of the CP-even Higgs bosons, $m_h$ and $\mH$.
As anticipated in the previous section, we take as input quantities 
the top pole mass $m_t = 175$ GeV, the electroweak symmetry--breaking
parameter $v = 246.218$ GeV, the OS masses $(\ma, \mtu, \mtd)$, the OS 
stop mixing angle $\thz$, and the $\ov{\rm DR}$ quantities $(\mu,\tan
\beta)$, evaluated at the reference scale $Q_0 = 175$ GeV. To facilitate
the comparison with the existing analyses of the two--loop corrected 
Higgs masses, we also refer to the unphysical parameters $(m_Q^{\rm OS}, 
m_U^{\rm OS}, X_t^{\rm OS})$ that can be derived by rotating the 
diagonal matrix of the OS stop masses by an angle $\thz$. However, it 
must be kept in mind that such parameters are `on-shell' only in the 
sense that they are extracted from the OS masses and mixing angle, but 
they do not correspond directly to any physical quantity. The gluino 
mass $\mgl$ and the left sbottom mass $m_{\widetilde{b}_L}$ enter 
only in the two--loop part of the calculation, thus their precise 
definition amounts to to a three--loop effect and is irrelevant to 
the perturbative order we are working at. Anyway, $m_{\widetilde{b}_L}$ 
is not really a free parameter, i.e. it has to be derived from the 
parameters of the stop sector: a simple choice consistent with our
approximations is $m_{\widetilde{b}_L} = m_Q^{\rm OS}$. Finally, the 
remaining input quantities are $\mz = 91.187$~GeV and the strong 
coupling constant $\as(\mz) = 0.118$.

Since we are interested here in pointing out the relevance of the
${\cal O}(\at^2)$ corrections with respect to those ${\cal O}(\at)$
and ${\cal O}(\at\as)$, we include in the one--loop part of the
calculation only the contribution of the top--stop sector of the MSSM,
as given in \cite{ab}. A general discussion of the precise
determination of the Higgs masses as functions of the whole set of
MSSM parameters will be presented elsewhere. For the two--loop ${\cal
O}(\at\as)$ corrections we use the simple analytical formulae of
\cite{DSZ}, equivalent to the results of the diagrammatic calculation
\cite{hollik} implemented in the Fortran code {\em FeynHiggs}
\cite{feynhiggs}, with the advantage of being simpler and more 
flexible for the choice of the renormalization scheme. For the ${\cal
O}(\at^2)$ corrections to the CP--even Higgs mass matrix, we use our
complete analytical formulae in the version appropriate for OS input
parameters. In the analysis of \cite{hollik}, some `leading
logarithmic' ${\cal O}(\at^2)$ corrections have been added to
$\left({\cal M}_S^2\right)_{22}$, proportional to powers of
$\ln\,(\msqu\msqd/m_t^4)$ and obtained by renormalization group
methods \cite{rge}. In the following we will compare our complete
${\cal O}(\at^2)$ result with the renormalization group result,
and we will point out that, for some choices of the SUSY parameters,
the logarithmic corrections amount only to a fraction of the full
ones.

\begin{figure}[p]
\begin{center}
\mbox{
\hspace{-.4cm}
\epsfig{figure=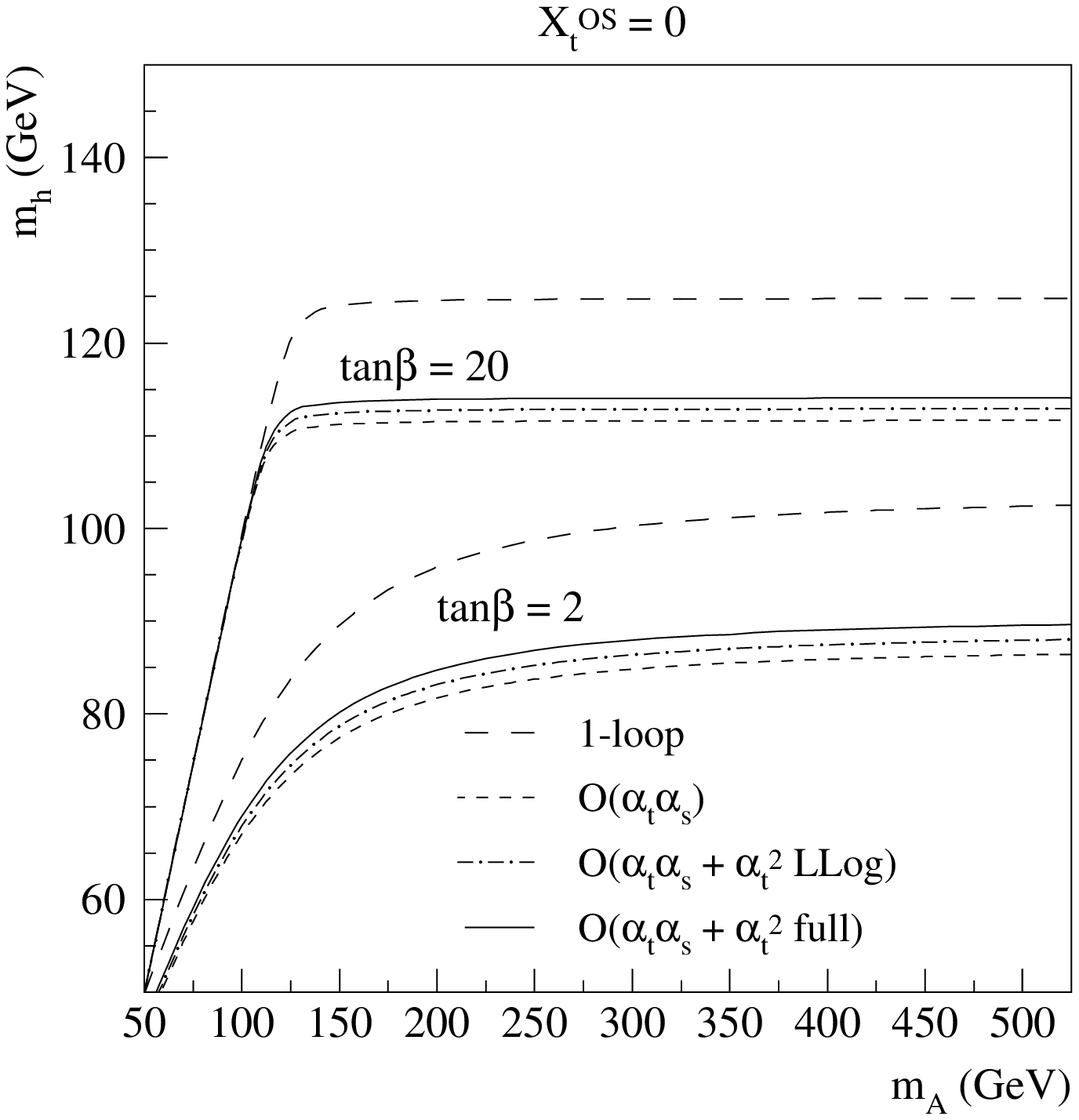,width=9.3cm,height=9.3cm}
\hspace{-1.5cm}
\epsfig{figure=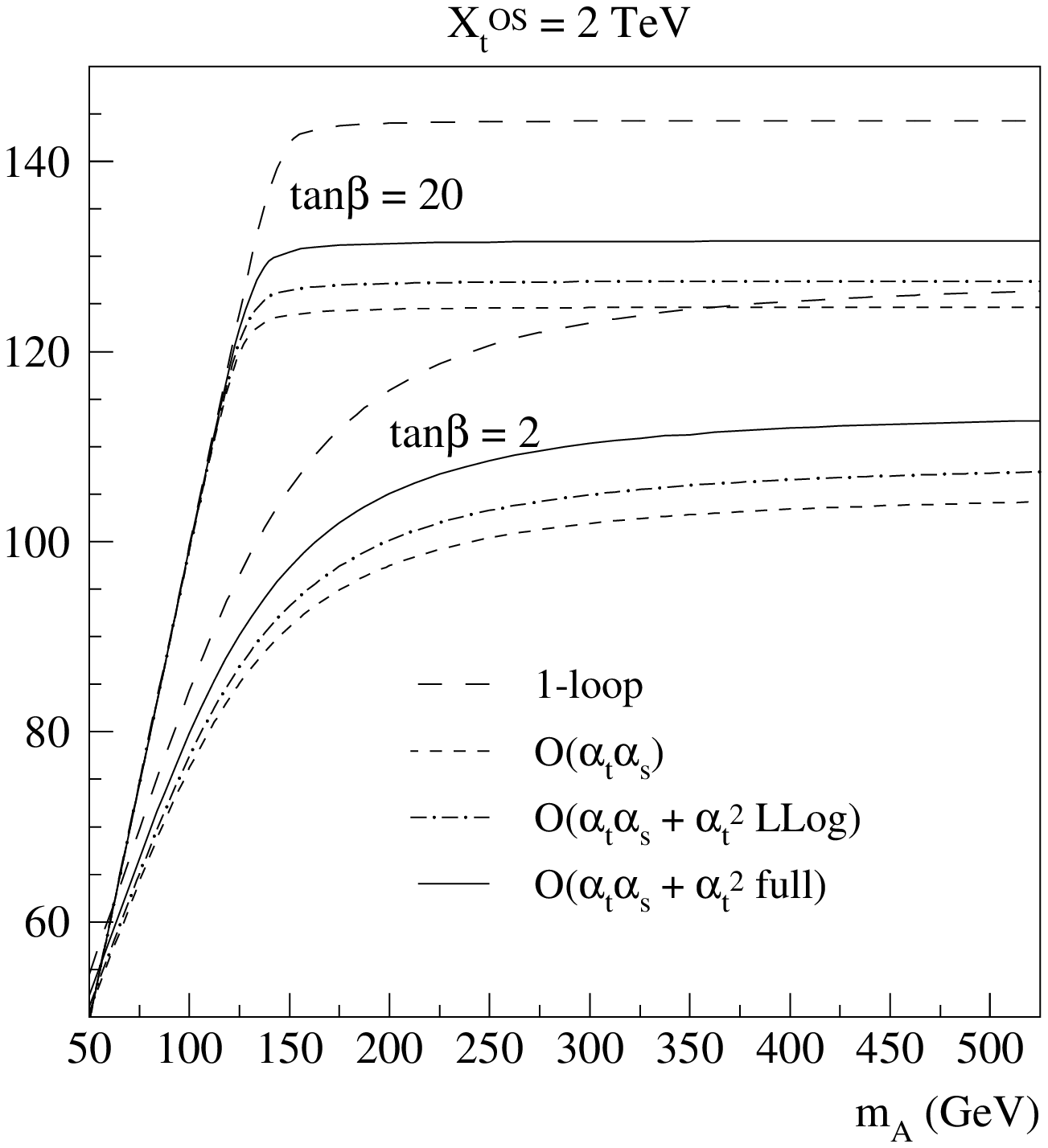,width=9.3cm,height=9.3cm}}
\end{center}
\vspace{-0.5cm}
\caption{The mass $m_h$ as a function of $\ma$,
for $\tan\beta = 2$ or 20 and $X_t^{\rm OS} = 0$ or 2 TeV.
The other parameters are $m_Q^{\rm OS} = m_U^{\rm OS} = 1$ TeV, 
$\mu = 200$ GeV, $\mgl = 800$ GeV. The meaning of the different
curves is explained in the text.}
\label{mhvsma}
\end{figure}
\begin{figure}[p]
\begin{center}
\mbox{
\hspace{-.4cm}
\epsfig{figure=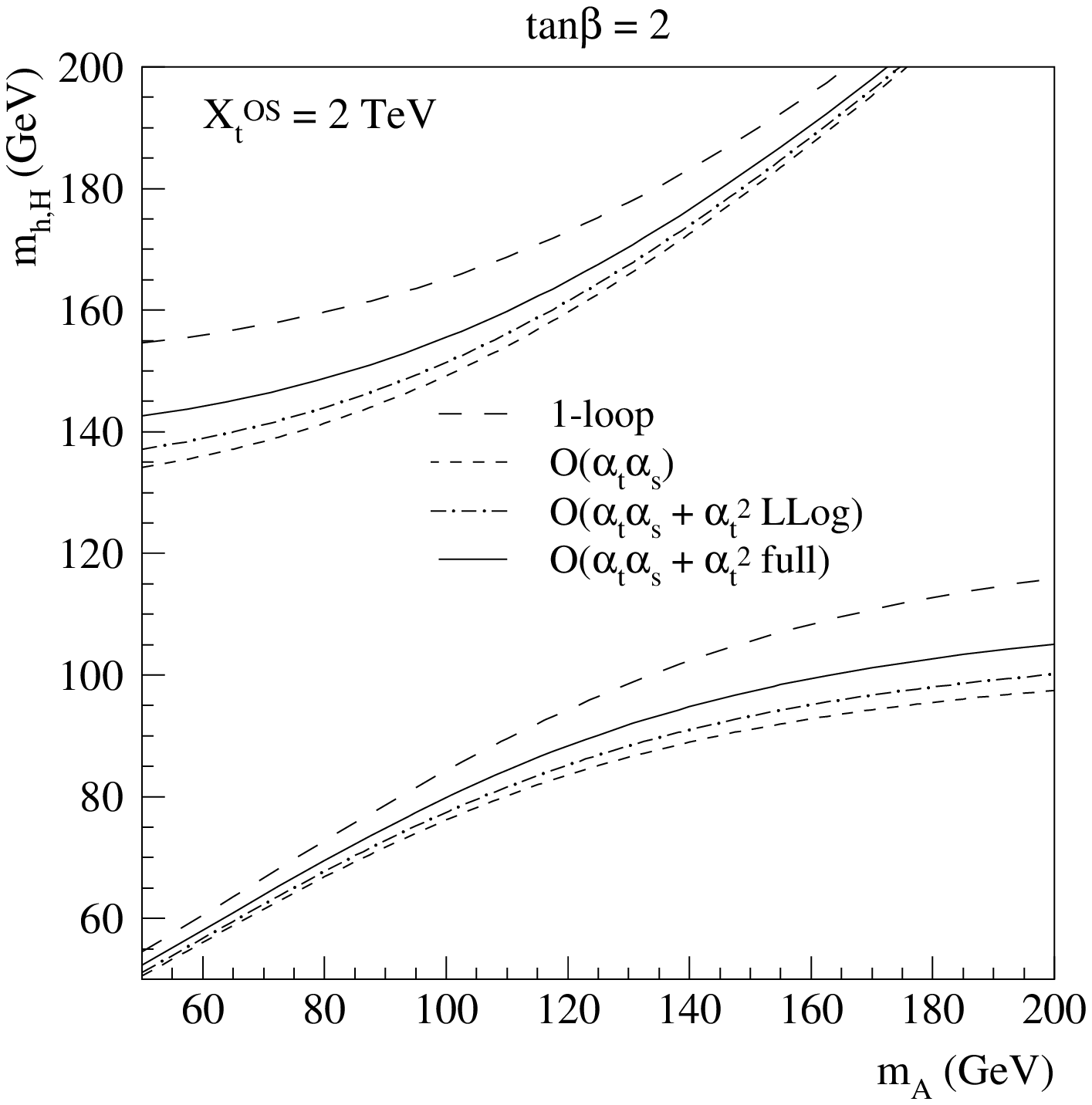,width=9.3cm,height=9.3cm}
\hspace{-1.5cm}
\epsfig{figure=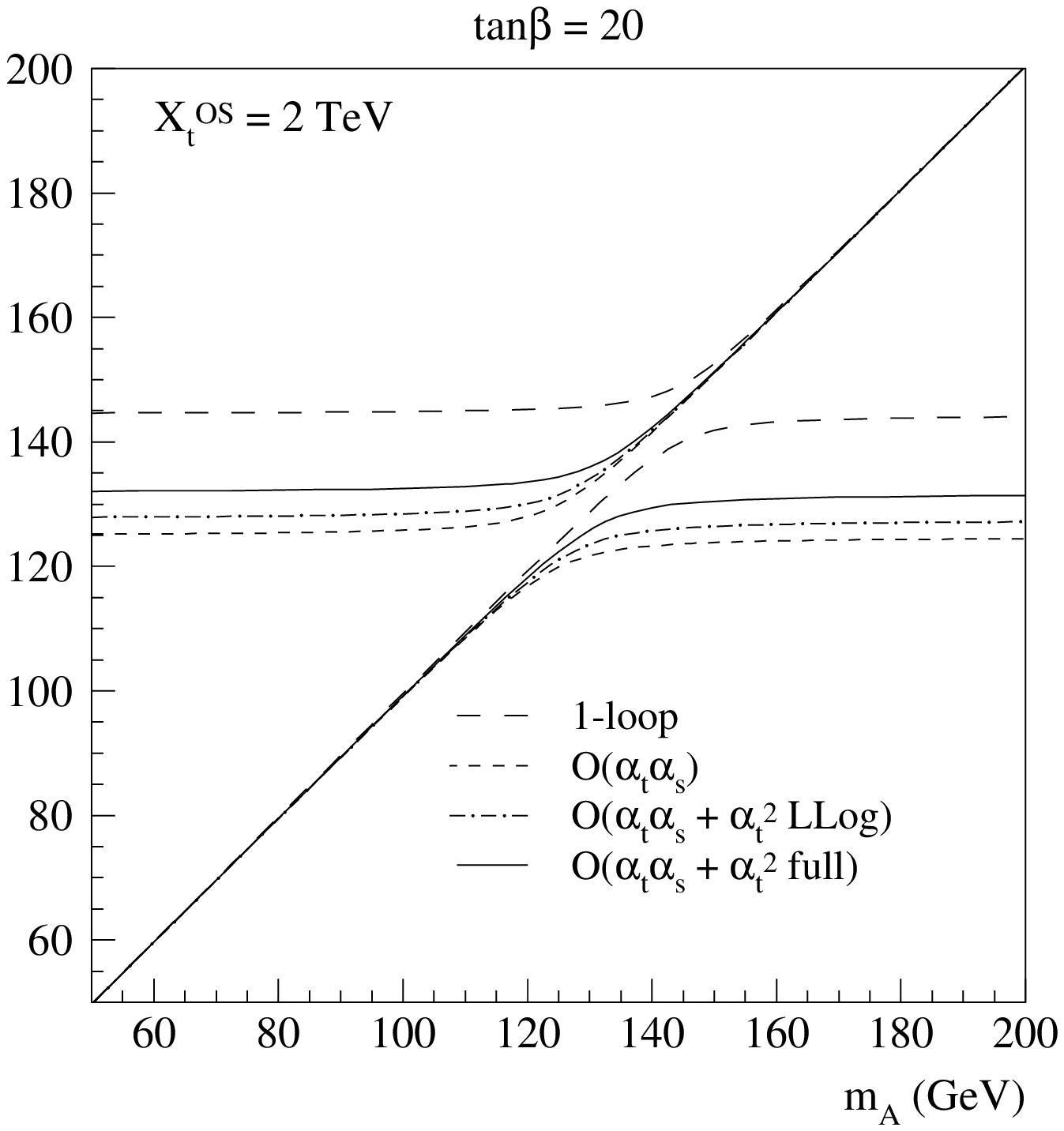,width=9.3cm,height=9.3cm}}
\end{center}
\vspace{-0.5cm}
\caption{The masses $(m_h,\mH)$ as functions
of $\ma$, for $\tan\beta = 2$ or 20 and $X_t^{\rm OS} = 2$ TeV.
The other input parameters are as in Fig.~\ref{mhvsma}.}
\label{mhhhvsma}
\end{figure}

Fig.~\ref{mhvsma} shows $m_h$ as a function of $\ma$, for $\tan
\beta = 2$ or 20 and $X_t^{\rm OS} = 0$ or 2 TeV. The other input 
parameters are chosen as $m_Q^{\rm OS} = m_U^{\rm OS} = 1$ TeV, 
$\mu = 200$ GeV, $\mgl = 800$ GeV. The cases with $X_t^{\rm OS} 
= 0$ and $X_t^{\rm OS} = 2$ TeV correspond, respectively, to the 
so-called `no--mixing' and `$m_h$--max' benchmark scenarios considered 
in the experimental analyses \cite{exp}. The values of the stop 
masses and mixing angle are approximately, neglecting small D--term 
contributions, $\mtu = \mtd = 1015$ GeV, $\thz = 0$ in the no--mixing 
case, and $\mtu = 1175$ GeV, $\mtd = 825$ GeV, $\thz = \frac{\pi}{4}$ 
in the case $X_t^{\rm OS} = 2$ TeV. The curves in Fig.~\ref{mhvsma} 
(and also in the other figures, unless indicated differently) 
correspond to the two--loop corrected Higgs mass at ${\cal O}(\at\as)$ 
(short--dashed line), at ${\cal O}(\at\as+\at^2)$ including only the 
logarithmic corrections of \cite{hollik} (dot--dashed line), and at 
${\cal O}(\at\as+\at^2)$ including our full computation (solid line). 
The one--loop result for $m_h$ is also shown for comparison (long--dashed 
line). It can be seen from the figure that the ${\cal O}(\at\as)$ 
corrections are in general large, reducing the one--loop result for
$m_h$ by 10--20 GeV. On the other hand, the ${\cal O}(\at^2)$ corrections 
tend to increase $m_h$: for small stop mixing they are generally small 
(less than 2--3 GeV), whereas for large stop mixing they can reach 7--8 
GeV, i.e. a non--negligible fraction of the ${\cal O}(\at\as)$ ones. 
Moreover, it appears that the `non--leading' corrections included
in our full result are always comparable in size with the `leading 
logarithmic' ones, and are even more significant than the latter in the 
case of large stop mixing. For $\ma \gg \mz$, we find good numerical 
agreement with the value of $m_h$ obtained from the approximate 
formula (20) of Ref.~\cite{ez2}.

In Fig.~\ref{mhhhvsma} we plot the one--loop and two--loop 
corrected values of both $m_h$ and $\mH$, in the region 
$\ma < 200$ GeV, for the case of large stop mixing. The other 
parameters are chosen as in Fig.~\ref{mhvsma}. It can be seen 
that the radiative corrections affect significantly $\mH$ only 
for quite low values of $\ma$ ($\ma \simlt 200$ GeV for small 
$\tan\beta$ and  $\ma \simlt 140$ GeV for large $\tan\beta$). 
We stress that the  determination of both mass eigenstates
is made possible by our implementation of the EPA, whereas
other existing two--loop calculations in the EPA \cite{hh,ez1,ez2} 
addressed only the corrections to $m_h$ in the limit $\ma \gg \mz$.

\begin{figure}[p]
\begin{center}
\mbox{
\hspace{-.4cm}
\epsfig{figure=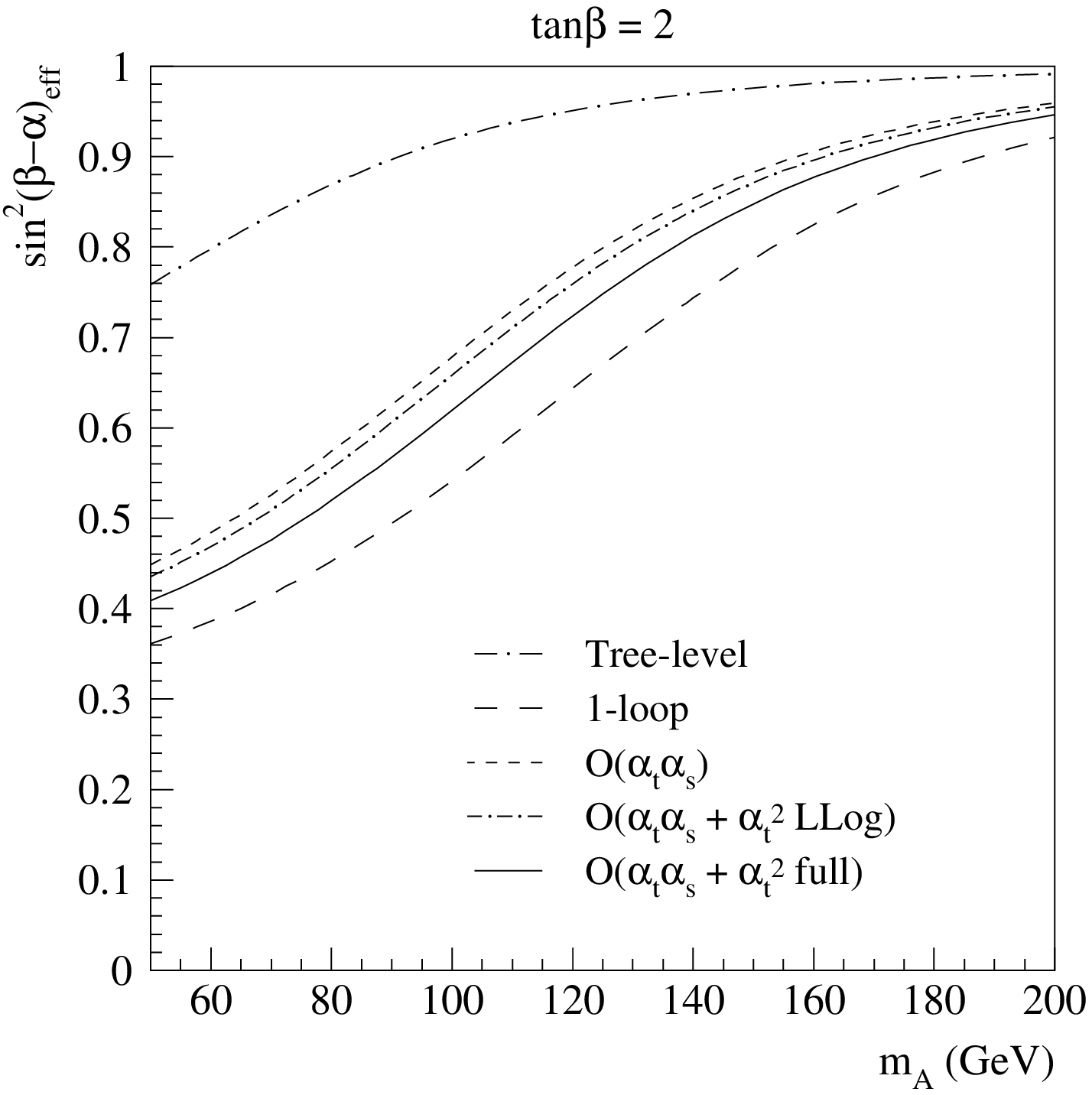,width=9.3cm,height=9.3cm}
\hspace{-1.5cm}
\epsfig{figure=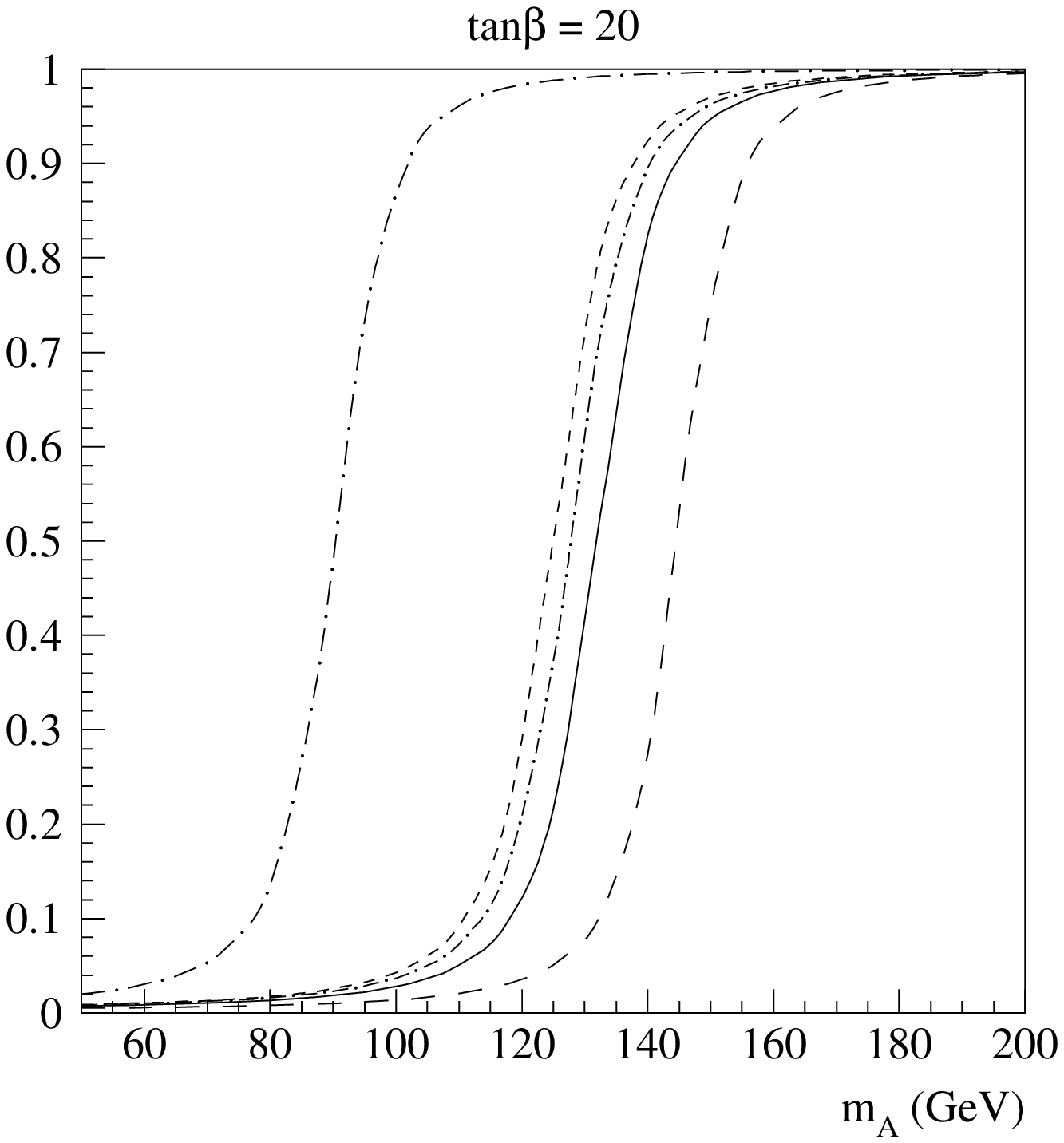,width=9.3cm,height=9.3cm}}
\end{center}
\vspace{-0.5cm}
\caption{The effective CP--even Higgs mixing angle $\ov{\alpha}$,
in the combination $\sin^2 (\beta - \ov{\alpha})$, as a function 
of $\ma$, for $\tan\beta = 2$ or 20 and 
$m_Q^{\rm OS} = m_U^{\rm OS} = 1$ TeV, $X_t^{\rm OS} = 2$ TeV, 
$\mu = 200$ GeV, $\mgl = 800$ GeV. The meaning of the curves,
explained in the text, is the same in the two frames.}
\label{alvsma}
\end{figure}
\begin{figure}[p]
\begin{center}
\mbox{
\hspace{-.4cm}
\epsfig{figure=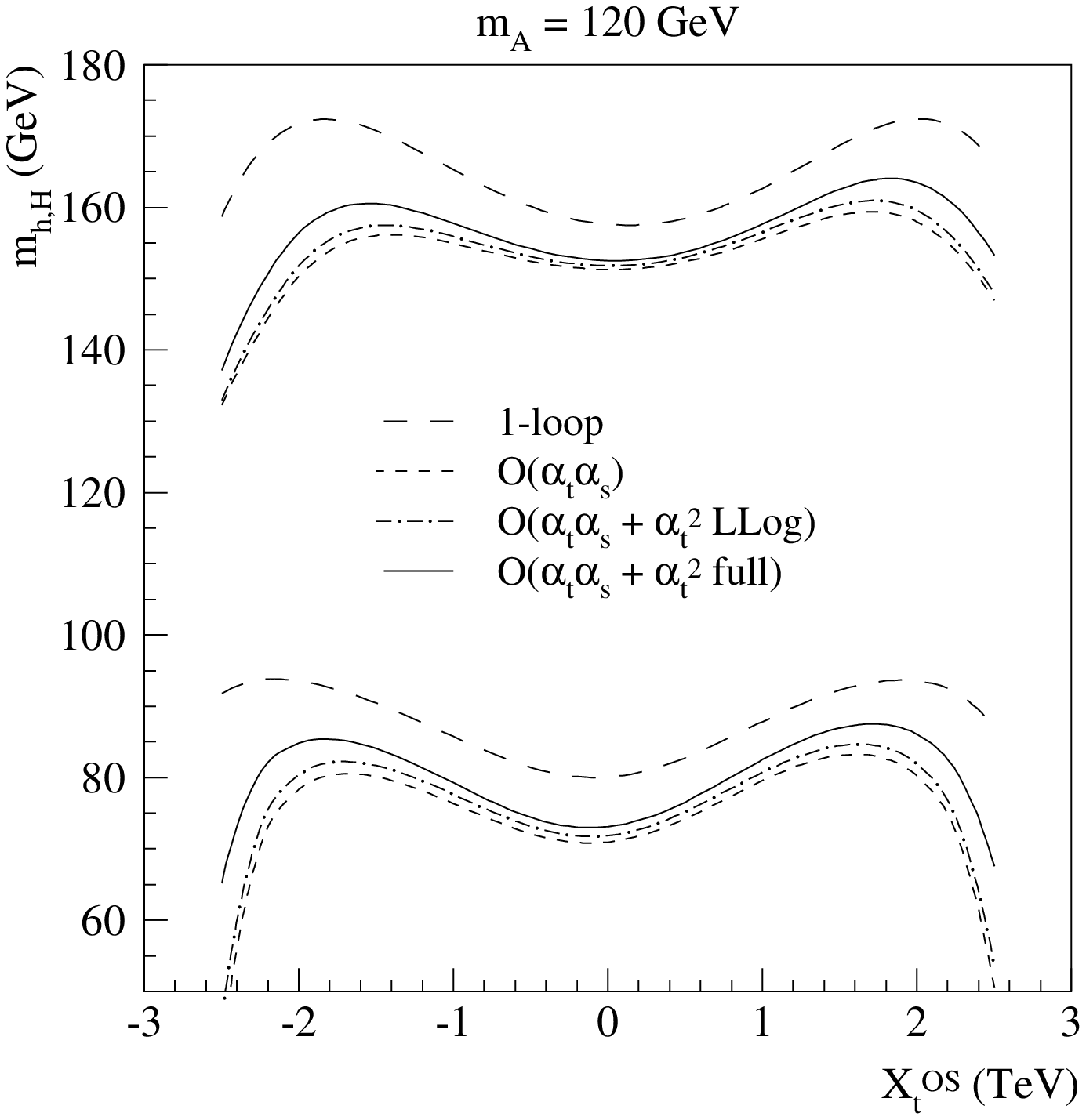,width=9.3cm,height=9.3cm}
\hspace{-1.5cm}
\epsfig{figure=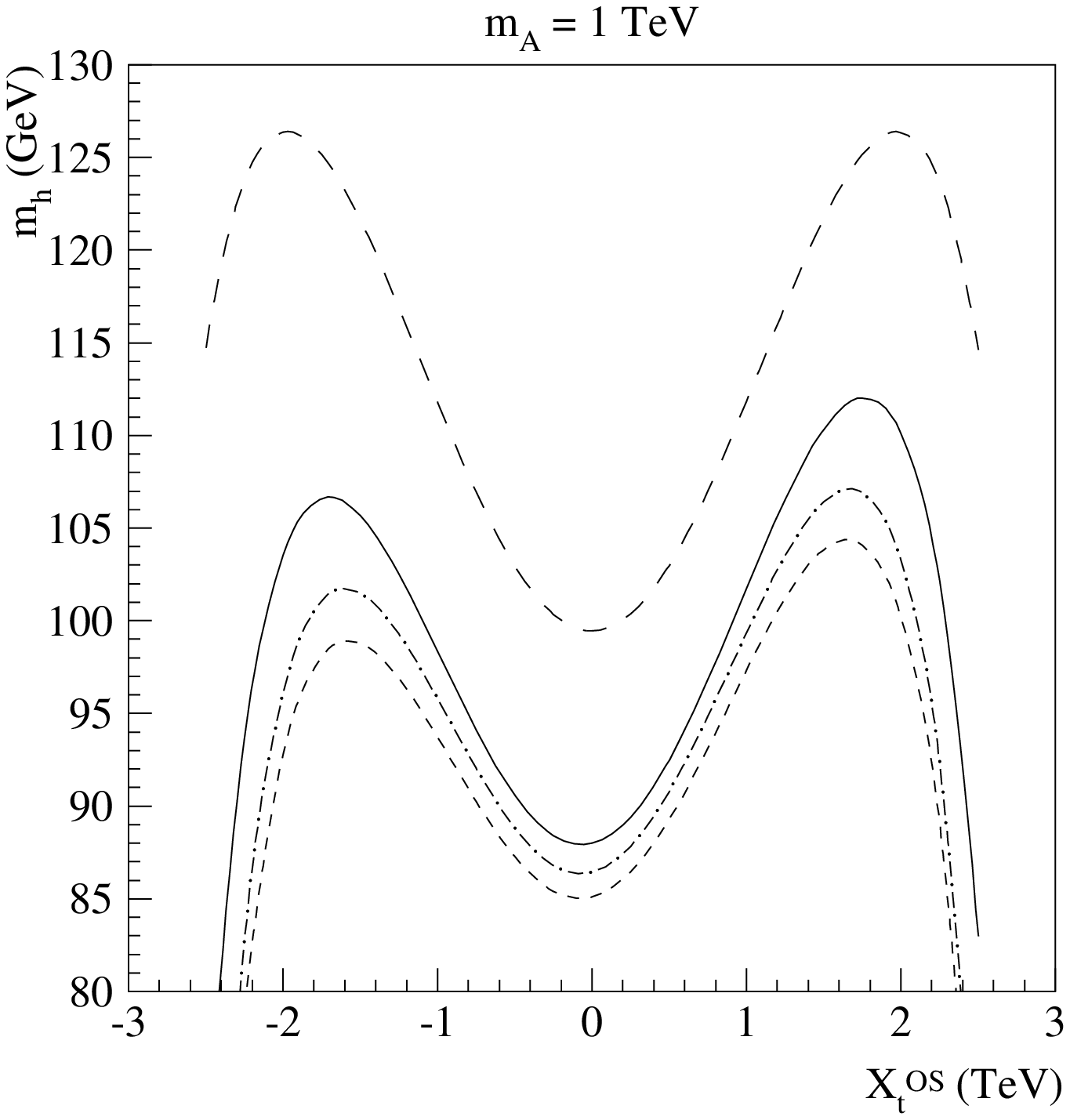,width=9.3cm,height=9.3cm}}
\end{center}
\vspace{-0.5cm}
\caption{The CP--even Higgs masses as functions of the stop mixing 
parameter $X_t^{OS}$, for $\ma = 120$ GeV or 1 TeV. 
For large $\ma$ only $m_h$ is shown. The other
input parameters are $m_Q^{\rm OS} = 1$ TeV, $m_U^{\rm OS} = 700$ GeV, 
$\tan\beta = 2$, $\mu = 200$ GeV, $\mgl = 800$ GeV.}
\label{mhvsx}
\end{figure}

Fig.~\ref{alvsma} shows the corrections to the effective CP--even
Higgs mixing angle $\ov{\alpha}$, for the case $X_t^{\rm OS} = 2$ TeV,
in the region of moderately low $\ma$ (for large $\ma$ the mixing
angle sticks to its tree-level value, $\beta - \frac{\pi}{2}$). The
other input parameters are as in Fig.~\ref{mhvsma}. We have chosen to
plot the combination $\sin^2(\beta-\ov{\alpha}\,)$ that enters the
cross-section for the process $e^+\,e^-\rightarrow h \,Z$. Again, it
turns out that the ${\cal O}(\at^2)$ corrections can reach 40\% of the
${\cal O}(\at\as)$ ones, and that the renormalization group method
provides only a fraction of the full two--loop Yukawa corrections. In
the interpretation of Fig.~\ref{alvsma}, however, the reader should
keep in mind that our renormalized angle $\ov{\alpha}$ does not
exhaust the ${\cal O} (\at \as + \at^2)$ corrections to the considered
process.

In Fig.~\ref{mhvsx} we explore in more detail the dependence of the
two--loop corrected Higgs masses on the stop mixing parameter,
$X_t^{\rm OS}$.  We set $\tan\beta = 2$ and $\ma = 120$ GeV or 1
TeV. Taking advantage of the fact that our analytical formulae are
valid for arbitrary values of the MSSM parameters, we split the values
of the diagonal entries of the stop mass matrix, choosing $m_Q^{\rm
OS} = 1$ TeV and $m_U^{\rm OS} = 700$ GeV (in fact the choice $m_Q =
m_U$, commonly considered for simplicity, seems to us quite unnatural,
since it is not preserved by potentially large ${\cal O}(\at)$
radiative corrections). Once more, it can be seen from
Fig.~\ref{mhvsx} that, for large stop mixing, the ${\cal O}(\at^2)$
corrections to $m_h$ (and also to $\mH$, for small $\ma$) amount to a
significant fraction of the ${\cal O}(\at\as)$ ones, and are not well
accounted for by the `leading logarithmic' terms.

\begin{figure}[t]
\begin{center}
\epsfig{figure=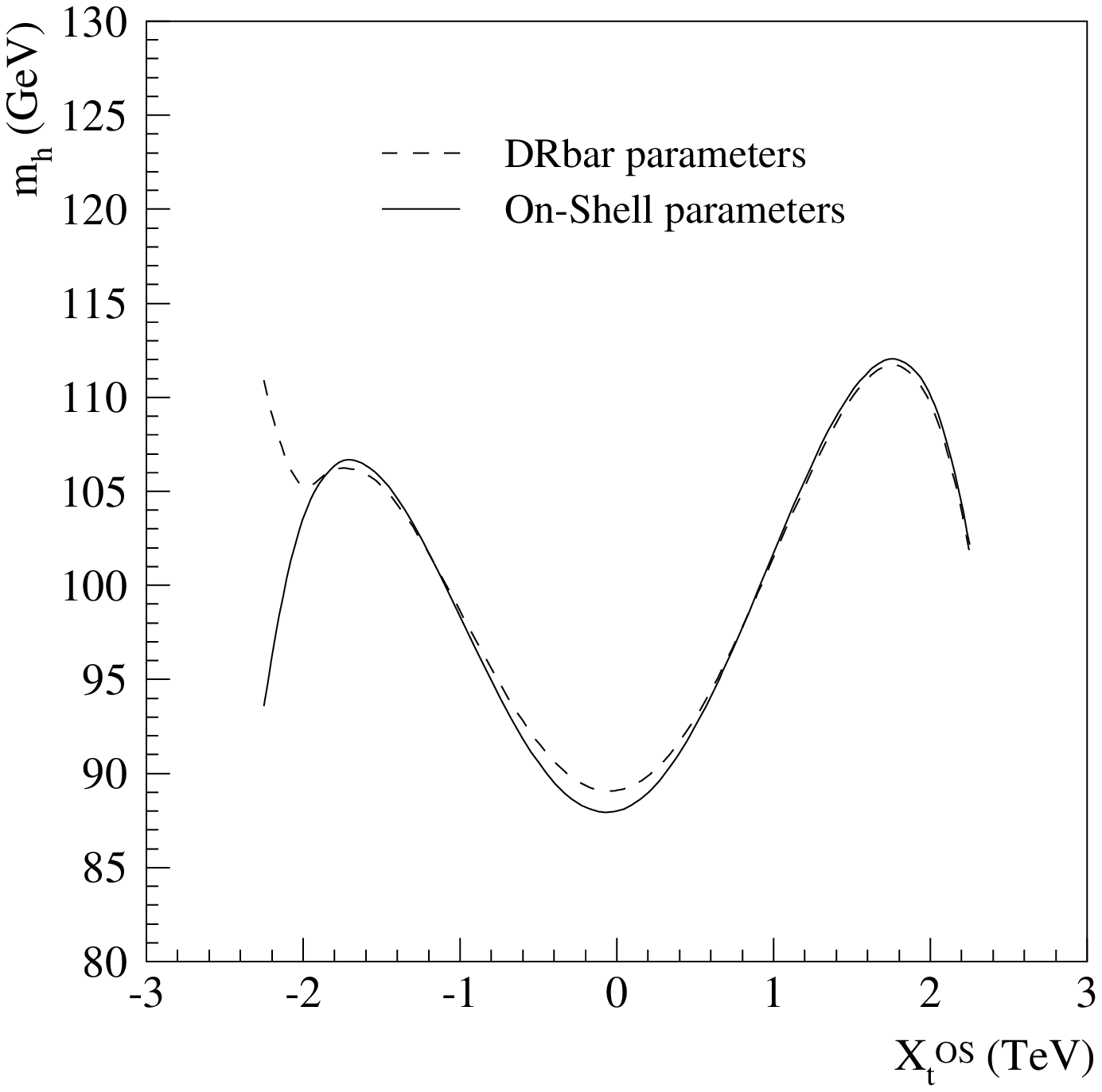,width=9.3cm,height=9.3cm}
\end{center}
\vspace{-0.5cm}
\caption{The mass $m_h$ as a function of the stop mixing 
parameter $X_t^{OS}$. The input parameters are chosen as in the right
frame of Fig.~\ref{mhvsx}. The solid line shows $m_h$ evaluated with 
OS parameters, whereas the dashed line shows $m_h$ evaluated with the 
corresponding $\ov{\rm DR}$ parameters.}
\label{schemes}
\end{figure}

Finally, in Fig.~\ref{schemes} we show the variation of our result for
$m_h$ under a change of renormalization scheme. We use the
same input parameters as in Fig.~\ref{mhvsx}, for the case $\ma =
1$~TeV. The solid line in Fig.~\ref{schemes} corresponds to $m_h$
evaluated with our choice of OS input parameters, as described in
Section~4. For comparison, we have converted the OS input parameters
into $\ov{\rm DR}$ quantities by means of
Eqs.~(\ref{dmtdv})--(\ref{detheta}), and we have used the latter as
input of the pure $\ov{\rm DR}$ formulae, using the original 
$\hat{F}_i^{2\ell}$
instead of the shifted $F_i^{2\ell}$ of Eqs.~(\ref{f12lr})--(\ref{dfam23}).
The results for $m_h$ are plotted as a dashed line. The differences
between the two curves can be interpreted in part as three--loop
effects, induced by the variation of the parameters that enter in the
two--loop formulae, and in part as two--loop ${\cal O}(g^2 \at)$
effects, due to the fact that the counterterms of
Eqs.~(\ref{f12lr})--(\ref{dfam23}) account only for the scheme
dependence of the ${\cal O}(\at)$ EPA corrections, whereas in the
one--loop corrections we include also the ${\cal O} (g^2)$ D--term
contributions and the momentum dependent parts of the propagators. The
figure shows that for moderate $X_t^{\rm OS}$ there is good agreement
between the two curves, the differences being ${\cal O}(1 \, {\rm
GeV})$ or smaller. On the other hand, for $X_t^{\rm OS}$ large and
negative the $\ov{\rm DR}$ calculation of $m_h$ blows up. The reason
is that, for such values of $X_t^{\rm OS}$, the stop masses and mixing
angle are subject to very large ${\cal O} (\at)$ corrections
proportional to $A_t^2$. As discussed in the previous section, a
similar phenomenon would also occur for a large gluino mass, because
of the non--decoupling properties of the $\ov{\rm DR}$ renormalization
scheme.
%
%%%%%%%%%%%%%%%%%%%%%%%%%%%%%%%%%%%%%%%%%%%%%%%%%%%%%%%%%%%%%%%%%%
%
\section{Conclusions}
\label{section:5}
In this paper we computed the ${\cal O}(\at^2)$ two--loop corrections
to the mass matrix of the neutral CP--even Higgs bosons in the MSSM,
extending earlier results \cite{hh,ez2} on $\mh$.  Using the formalism
of the effective potential, we obtained complete analytic expressions
for the momentum--independent part of these corrections, valid for
arbitrary values of $\ma$ and of the parameters in the stop
sector. The inclusion of momentum--dependent effects was explained in
Section~2. Our results can be efficiently inserted in general computer
codes, to obtain ${\cal O}(\at \as + \at^2)$ corrected expressions for
the masses $(m_h,\mH)$.
 
Our general analytic expressions are too long to be useful if
explicitly written on paper, thus we decided to make them available as
a Fortran or Mathematica package. As an illustration, we wrote down
explicit analytical formulae for the simplified case where $m_Q=m_U
\equiv M_S$ and ${\cal O}(m_t/M_S)$ corrections are neglected, but
$\ma$ and the stop mixing parameters are left arbitrary.

We have presented our results in such a way that the input parameters
of the top and stop sectors can be assigned either in the $\ov{\rm
DR}$ scheme or in any other renormalization scheme, in particular
those with on--shell definitions of the masses and of the stop mixing
angle. We have also included in our computer package the analytic
result for the ${\cal O}(\at^2)$ two--loop corrected relation between
$m_3^2$, defined in the $\ov{\rm DR}$ scheme, and the EPA mass
$\mabar^2$, with the remaining parameters assigned in the $\ov{\rm
DR}$ scheme.

As discussed in Section~\ref{section:4}, the ${\cal O} (\at^2)$
corrections have a significant impact on the predictions for $m_h$ and
$\mH$. While the ${\cal O} (\at\as)$ corrections typically decrease
the values of $(m_h,\mH)$, the ${\cal O} (\at^2)$ ones typically
increase them, producing a partial compensation: in the case of large
stop mixing, the latter can reach up to $40\, \%$ of the former, and
shift $\mh$ upwards by 7--8~GeV. Also, our general result for the
${\cal O} (\at^2)$ corrections, both to the masses $(m_h,\mH)$ and to
the mixing angle $\ov{\alpha}$, is quite different from the one
obtained via a `leading logarithm' renormalization group analysis: the
shift in $m_h$ is twice as large for small stop mixing, about three
times larger for large stop mixing.

The present paper and its companion, Ref.~\cite{DSZ}, contain the most
advanced calculation (so far) of the two--loop corrections to the MSSM
neutral Higgs boson masses. The ${\cal O}(\at \as + \at^2)$
corrections to $\mH^2$ are still affected by some residual
uncertainties for large $\ma$, because of momentum--dependent effects
not computed so far.  However, the relative importance of such
corrections is small, since the tree--level value of $\mH$ grows with
$\ma$. As explained in Section~2, no such uncertainties exist for
$m_h$. The main residual two--loop corrections to the Higgs masses are
controlled by the electroweak gauge couplings and by the bottom Yukawa
coupling. In particular, two--loop corrections proportional to $\ab$
are expected to be numerically non--negligible only for $\tan \beta
\simgt 40$, i.e. when $\ab$ is comparable with $\at$, despite the fact
that $m_b \ll \mt$.  The ${\cal O}(\ab \as m_b^2)$ and ${\cal O}(
\ab^2 m_b^2)$ corrections could be computed by performing simple 
substitutions in the formulae for the top case,
whereas the ${\cal O}(\at \ab m_b^2)$ and ${\cal O}(\at \ab m_t^2)$
corrections require further work. We also recall that the bottom mass
$m_b$ can receive quite large threshold corrections \cite{hrs} that
could induce additional complications.

In conclusion, our work should lead to a more accurate interpretation
of the experimental searches for the neutral MSSM Higgs bosons at LEP,
the Tevatron, the LHC and other possible future colliders. The
importance of the new ${\cal O}(\at^2)$ effects we have computed will
increase further when the top quark mass is measured more precisely
(we recall that the present CDF/D0 average~\cite{top} is $m_t = 174.3
\pm 5.1$~GeV). Then, we can hope for the next step, the discovery of
supersymmetric particles and supersymmetric Higgs bosons.
%
%%%%%%%%%%%%%%%%%%%%%%%%%%%%%%%%%%%%%%%%%%%%%%%%%%%%%%%%%%%%%%%%%%
%
\section*{Acknowledgements}
One of us (F.Z.) would like to thank the Physics Department of the
University of Padua for its hospitality during part of this project,
and INFN, Sezione di Padova, for partial travel support. This work
was supported in part by the European Union under the contracts
HPRN-CT-2000-00149 (Collider Physics) and HPRN-CT-2000-00148 (Across
the Energy Frontier).
%
%%%%%%%%%%%%%%%%%%%%%%%%%%%%%%%%%%%%%%%%%%%%%%%%%%%%%%%%%%%%%%%%%%
%
\begin{appendletterA}
\section*{Appendix: One-loop self-energies}

We provide here the explicit formulae for the one-loop self-energies of 
the top quark and squarks and of the W-boson. As discussed in Section 4, 
these formulae are required to relate the   
parameters $m_t\,,\,\msqu\,,\,\msqd\,,\,\s2t$ and $v\,$
defined in the $\ov{\rm DR}$ scheme to the corresponding 
quantities chosen in that Section. Since we are only interested in the 
terms that generate 
${\cal O}(\at^2)$ corrections at two-loops, we neglect all gauge couplings 
and we keep $h_t$ as the only non-vanishing Yukawa coupling. This leads
to the simplified mass spectrum described at the beginning of Section 3.
The real finite parts of the self-energies for the top and the stops are: 

\bea
\hat{\Sigma}_t\,(\mt) &= &\frac{h_t^2}{32\,\pi^2}\,\mt\,\left\{\;
 \left(1- \frac52\sb^2  \right) \ln \frac{\mt^2}{Q^2} 
+ \frac{3\,\cb^2}2 \,\frac{\ma^2}{\mt^2} 
\left( 1 - \ln \frac{\ma^2}{Q^2} \right) 
-\frac32 \frac{\mu^2}{\mt^2} \left( 1 - \ln \frac{\mu^2}{Q^2} \right)
\right.\nn\\
&&\nn\\
&&+ 5\,\sb^2 -1 + 
\frac{\cb^2}2 \left( 1 -\frac{\ma^2}{\mt^2} \right)\, \hat{B}_0
(\mt^2,0,\ma^2) + \cb^2\,\left( 2 -\frac{\ma^2}{\mt^2} \right) 
\hat{B}_0(\mt^2,\mt^2,\ma^2)\nn\\
&&\nn\\
&&+ 
\frac12 \left[ \frac{\msqu}{\mt^2} \left( 1 - \ln \frac{\msqu}{Q^2} \right)
 + \frac{\mt^2 - \msqu +\mu^2}{\mt^2} \hat{B}_0(\mt^2,\mu^2,\msqu) \right.\nn\\
&& \left.\left. \phantom{\left( 1 - \ln \frac{\msqu}{Q^2} \right)}
\hspace{3.5cm}\; +\; (\tilde{t}_1 \rightarrow \tilde{t}_2) 
\;+ \;(\tilde{t}_1 \rightarrow \tilde{b}_L)\,
\right] \,\right\}\,, \label{eq:a1}\\[0.3cm]
\hat{\Pi}_{11}\,(\msqu) & = & \frac{h_t^2}{16\,\pi^2} \,\left\{\;
\cb^2\,(1+\st^2) \,A_0\,(\ma^2) - A_0(\mt^2) - (1+\st^2)\,A_0(\mu^2)   
\right.\nn\\
&&\nn\\
&&+ \st^2\,A_0\,(\mbl) 
+ \left( \c2t^2 - \frac{N_c-1}{2}\,\s2t^2\right)\,A_0(\msqd)
+ \frac{N_c+1}{2}\,\s2t^2\,A_0\,(\msqu)\nn\\
&&\nn\\
&&+(\msqu -\mt^2 -\mu^2)\,
\hat{B}_0(\msqu,\mt^2,\mu^2) + \st^2 \, (\msqu-\mu^2)
\hat{B}_0(\msqu,0,\mu^2)\nn\\
&&\nn\\
&&+\hlf \,\left[\sb^2\,(2\,\mt + \s2t\,X_t)^2\,\hat{B}_0(\msqu,\msqu,0)
+ \cb^2\,(2\,\mt +  \s2t\,Y_t)^2\,\hat{B}_0(\msqu,\msqu,\ma^2)\right.\nn\\
&&\nn\\
&&+\left.
\sb^2\,(1 + \c2t^2)\,X_t^2\,\hat{B}_0(\msqu,\msqd,0)
+\cb^2\,(1 + \c2t^2)\,Y_t^2\,\hat{B}_0(\msqu,\msqd,\ma^2)\;\right]\nn\\
&&\nn\\
&&+ \left. \sb^2\,(\mt\,\ct + X_t\,\st)^2\,\hat{B}_0(\msqu,\mbl,0)
+ \cb^2\,(\mt\,\ct + Y_t\,\st)^2\,\hat{B}_0(\msqu,\mbl,\ma^2)\;\right\}
\,,\nn\\
&&\label{eq:a2}\\[0.3cm]
&&\nn\\
\hat{\Pi}_{12}(p^2) & = & \frac{h_t^2}{32\,\pi^2} \,\left\{\;
\s2t \,( p^2 - \mu^2)\,\hat{B}_0(p^2,0,\mu^2) 
-\s2t \,A_0(\mu^2)\right.\nn\\
&&\nn\\
&&+ \s2t\,\cb^2\,A_0\,(\ma^2) + \s2t\,A_0\,(\mbl) 
+ (N_c+1)\,\c2t\,\s2t\,\left[ A_0\,(\msqu) - A_0\,(\msqd)\right]\nn\\
&&\nn\\
&&+ \sb^2\,\c2t\,X_t\,(2\,\mt + \s2t\,X_t)\,\hat{B}_0(p^2,\msqu,0)
+ \cb^2\,\c2t\,Y_t\,(2\,\mt + \s2t\,Y_t)\,\hat{B}_0(p^2,\msqu,\ma^2)\nn\\
&&\nn\\
&&+ \sb^2\,\c2t\,X_t\,(2\,\mt - \s2t\,X_t)\,\hat{B}_0(p^2,\msqd,0)
+ \cb^2\,\c2t\,Y_t\,(2\,\mt - \s2t\,Y_t)\,\hat{B}_0(p^2,\msqd,\ma^2)\nn\\
&&\nn\\
&&- 2\,\sb^2\,(\mt\,\ct + X_t \,\st)\,(\mt\,\st - X_t \,\ct)
\,\hat{B}_0(p^2,\mbl,0)\nn\\
&&\nn\\
&&- \left. 2\,\cb^2\,(\mt\,\ct + Y_t \,\st)\,(\mt\,\st - Y_t \,\ct)
\,\hat{B}_0(p^2,\mbl,\ma^2)\;\right\}\,,
\eea
where 
\be
A_0(m^2) = m^2 \left(1 - \ln \frac{m^2}{Q^2} \right)\,, \nn\\
\ee
and $\hat{B}_0$ denotes the real finite 
part of the $B_0$  Passarino-Veltman function, i.e.~:
\be
\hat{B}_0\,(p^2,m_1^2,m_2^2)  =  
- {\rm Re} \int^1_0 dx \ln 
\frac{(1-x)\, m_1^2 + x\,m_2^2 - x\,(1-x)\,p^2 - i\epsilon}{Q^2} \, .
\ee
An explicit expression for $B_0$ can be found e.g. in \cite{DS}.
The function $\hat{\Pi}_{22}\,(\msqd)$ can be obtained from the 
right hand side of Eq.~(\ref{eq:a2}) with the replacements $\msqu 
\leftrightarrow \msqd\,,\; \ct \rightarrow \st\,,\; \st \rightarrow 
-\ct$ (also implying $\s2t \rightarrow -\s2t\,, \c2t \rightarrow -\c2t$).

Finally the identification  $v = (\sq2\, G_{\mu})^{-1/2} = 246.218$ GeV
requires the self-energy of the $W$ boson at zero momentum transfer.
Including only the contributions from the top--bottom sector relevant
for our calculation:
\bea
\hat{\Pi}^{\;T}_{WW}(0) &=& \frac{g^2}{16\,\pi^2} N_c
\left\{ \mt^2\, \left(-\frac14 + \frac12 \ln \frac{\mt^2}{Q^2} \right)
     - \ct^2 
\left(
\frac{\msqu + \mbl}4  - \frac{\msqu \, \mbl}{2\,(\msqu-\mbl)} 
\ln \frac{\msqu}{\mbl} \right) \right. \nn\\
&&\nn\\
&& \hspace{2cm} - \st^2 \left.
\left(
\frac{\msqd + \mbl}4  - \frac{\msqd \, \mbl}{2\,(\msqd-\mbl)} 
\ln \frac{\msqd}{\mbl} \right) \right\}~~.
\eea
\end{appendletterA}
%
%%%%%%%%%%%%%%%%%%%%%%%%%%%%%%%%%%%%%%%%%%%%%%%%%%%%%%%%%%%%%%%%%%
%

\end{document}